%% file: main.tex
\documentclass{article}

\usepackage{t1enc} % \k
\usepackage{amsmath}
\usepackage{latexsym}
\usepackage{theorem}
\usepackage{amssymb}
\usepackage{url}

\input{prooftree}
\input{macros}

\input{macros2}

\begin{document}

\input{title}

\input{intro}
\input{cac}
\input{sub}
\input{typ}
\input{sn}
\input{cons}

\input{expl}

\input{prf}
\input{conclu}

% biblio
%\bibliographystyle{plain}
%\bibliography{abbrev,long-names,mybib}

\input{biblio}
% annexes
\input{trans}

\input{exp}

% end document
\end{document}

%% file: prooftree.tex
\newdimen\proofrulebreadth \proofrulebreadth=.05em
\newdimen\proofdotseparation \proofdotseparation=1.25ex
\newdimen\proofrulebaseline \proofrulebaseline=2ex
\newcount\proofdotnumber \proofdotnumber=3
\let\then\relax
\def\hfi{\hskip0pt plus.0001fil}
\mathchardef\squigto="3A3B
\newif\ifinsideprooftree\insideprooftreefalse
\newif\ifonleftofproofrule\onleftofproofrulefalse
\newif\ifproofdots\proofdotsfalse
\newif\ifdoubleproof\doubleprooffalse
\let\wereinproofbit\relax
\newdimen\shortenproofleft
\newdimen\shortenproofright
\newdimen\proofbelowshift
\newbox\proofabove
\newbox\proofbelow
\newbox\proofrulename
\def\shiftproofbelow{\let\next\relax\afterassignment\setshiftproofbelow\dimen0 }
\def\shiftproofbelowneg{\def\next{\multiply\dimen0 by-1 }%
\afterassignment\setshiftproofbelow\dimen0 }
\def\setshiftproofbelow{\next\proofbelowshift=\dimen0 }
\def\setproofrulebreadth{\proofrulebreadth}

\def\prooftree{% NESTED ZERO (\ifonleftofproofrule)
\ifnum  \lastpenalty=1
\then   \unpenalty
\else   \onleftofproofrulefalse
\fi
\ifonleftofproofrule
\else   \ifinsideprooftree
        \then   \hskip.5em plus1fil
        \fi
\fi
\bgroup% NESTED ONE (\proofbelow, \proofrulename, \proofabove,
\setbox\proofbelow=\hbox{}\setbox\proofrulename=\hbox{}%
\let\justifies\proofover\let\leadsto\proofoverdots\let\Justifies\proofoverdbl
\let\using\proofusing\let\[\prooftree
\ifinsideprooftree\let\]\endprooftree\fi
\proofdotsfalse\doubleprooffalse
\let\thickness\setproofrulebreadth
\let\shiftright\shiftproofbelow \let\shift\shiftproofbelow
\let\shiftleft\shiftproofbelowneg
\let\ifwasinsideprooftree\ifinsideprooftree
\insideprooftreetrue
\setbox\proofabove=\hbox\bgroup$\displaystyle % NESTED TWO
\let\wereinproofbit\prooftree
\shortenproofleft=0pt \shortenproofright=0pt \proofbelowshift=0pt
\onleftofproofruletrue\penalty1
}

\def\eproofbit{% NESTED TWO
\ifx    \wereinproofbit\prooftree
\then   \ifcase \lastpenalty
        \then   \shortenproofright=0pt  % 0: some other object, no indentation
        \or     \unpenalty\hfil         % 1: empty hypotheses, just glue
        \or     \unpenalty\unskip       % 2: just had a tree, remove glue
        \else   \shortenproofright=0pt  % eh?
        \fi
\fi
\global\dimen0=\shortenproofleft
\global\dimen1=\shortenproofright
\global\dimen2=\proofrulebreadth
\global\dimen3=\proofbelowshift
\global\dimen4=\proofdotseparation
\global\count255=\proofdotnumber
$\egroup  % NESTED ONE
\shortenproofleft=\dimen0
\shortenproofright=\dimen1
\proofrulebreadth=\dimen2
\proofbelowshift=\dimen3
\proofdotseparation=\dimen4
\proofdotnumber=\count255
}

\def\proofover{% NESTED TWO
\eproofbit % NESTED ONE
\setbox\proofbelow=\hbox\bgroup % NESTED TWO
\let\wereinproofbit\proofover
$\displaystyle
}%
\def\proofoverdbl{% NESTED TWO
\eproofbit % NESTED ONE
\doubleprooftrue
\setbox\proofbelow=\hbox\bgroup % NESTED TWO
\let\wereinproofbit\proofoverdbl
$\displaystyle
}%
\def\proofoverdots{% NESTED TWO
\eproofbit % NESTED ONE
\proofdotstrue
\setbox\proofbelow=\hbox\bgroup % NESTED TWO
\let\wereinproofbit\proofoverdots
$\displaystyle
}%
\def\proofusing{% NESTED TWO
\eproofbit % NESTED ONE
\setbox\proofrulename=\hbox\bgroup % NESTED TWO
\let\wereinproofbit\proofusing
\kern0.3em$
}

\def\endprooftree{% NESTED TWO
\eproofbit % NESTED ONE
  \dimen5 =0pt% spread of hypotheses
\dimen0=\wd\proofabove \advance\dimen0-\shortenproofleft
\advance\dimen0-\shortenproofright
\dimen1=.5\dimen0 \advance\dimen1-.5\wd\proofbelow
\dimen4=\dimen1
\advance\dimen1\proofbelowshift \advance\dimen4-\proofbelowshift
\ifdim  \dimen1<0pt
\then   \advance\shortenproofleft\dimen1
        \advance\dimen0-\dimen1
        \dimen1=0pt
        \ifdim  \shortenproofleft<0pt
        \then   \setbox\proofabove=\hbox{%
                        \kern-\shortenproofleft\unhbox\proofabove}%
                \shortenproofleft=0pt
        \fi
\fi
\ifdim  \dimen4<0pt
\then   \advance\shortenproofright\dimen4
        \advance\dimen0-\dimen4
        \dimen4=0pt
\fi
\ifdim  \shortenproofright<\wd\proofrulename
\then   \shortenproofright=\wd\proofrulename
\fi
\dimen2=\shortenproofleft \advance\dimen2 by\dimen1
\dimen3=\shortenproofright\advance\dimen3 by\dimen4
\ifproofdots
\then
        \dimen6=\shortenproofleft \advance\dimen6 .5\dimen0
        \setbox1=\vbox to\proofdotseparation{\vss\hbox{$\cdot$}\vss}%
        \setbox0=\hbox{%
                \advance\dimen6-.5\wd1
                \kern\dimen6
                $\vcenter to\proofdotnumber\proofdotseparation
                        {\leaders\box1\vfill}$%
                \unhbox\proofrulename}%
\else   \dimen6=\fontdimen22\the\textfont2 % height of maths axis
        \dimen7=\dimen6
        \advance\dimen6by.5\proofrulebreadth
        \advance\dimen7by-.5\proofrulebreadth
        \setbox0=\hbox{%
                \kern\shortenproofleft
                \ifdoubleproof
                \then   \hbox to\dimen0{%
                        $\mathsurround0pt\mathord=\mkern-6mu%
                        \cleaders\hbox{$\mkern-2mu=\mkern-2mu$}\hfill
                        \mkern-6mu\mathord=$}%
                \else   \vrule height\dimen6 depth-\dimen7 width\dimen0
                \fi
                \unhbox\proofrulename}%
        \ht0=\dimen6 \dp0=-\dimen7
\fi
\let\doll\relax
\ifwasinsideprooftree
\then   \let\VBOX\vbox
\else   \ifmmode\else$\let\doll=$\fi
        \let\VBOX\vcenter
\fi
\VBOX   {\baselineskip\proofrulebaseline \lineskip.2ex
        \expandafter\lineskiplimit\ifproofdots0ex\else-0.6ex\fi
        \hbox   spread\dimen5   {\hfi\unhbox\proofabove\hfi}%
        \hbox{\box0}%
        \hbox   {\kern\dimen2 \box\proofbelow}}\doll%
\global\dimen2=\dimen2
\global\dimen3=\dimen3
\egroup % NESTED ZERO
\ifonleftofproofrule
\then   \shortenproofleft=\dimen2
\fi
\shortenproofright=\dimen3
\onleftofproofrulefalse
\ifinsideprooftree
\then   \hskip.5em plus 1fil \penalty2
\fi
}

%% file: macros.tex
\newenvironment{ded}
   {\begin{center}\begin{prooftree}}
   {\end{prooftree}\end{center}}

\newcommand{\comment}[1]{}
\newcommand{\bu}{$\bullet$}
\newcommand{\hs}[1][3ex]{\hspace*{#1}}

\newcommand{\moins}{\setminus}
\newcommand{\vide}{\emptyset}

\newcommand{\xu}{\{x\to u\}}

\newcommand{\xS}{\{x\to S\}}

\newcommand{\vyu}{\{\vy\to\vu\}}

\newcommand{\vxl}{\{\vx\to\vl\}}

\newcommand{\dom}{\mr{dom}}

\newcommand{\FV}{\mr{FV}}
\newcommand{\pos}{\mr{Pos}}

\renewcommand{\a}{\rightarrow}
\newcommand{\A}{\Rightarrow}

\newcommand{\ad}{\downarrow}

\renewcommand{\to}{\mapsto}

\newcommand{\ab}{\a_\b}

\newcommand{\als}[1][]{~_{#1}^*\!\!\leftarrow}

\newcommand{\I}[1]{[\![#1]\!]}

\newcommand{\all}{\forall}

\newcommand{\bigou}{\bigvee}
\newcommand{\biget}{\bigwedge}

\newcommand{\st}{\star}
\newcommand{\B}{\Box} % latexsym
\renewcommand{\th}{\vdash}

\newcommand{\sle}{\subseteq}

\newcommand{\tgt}{\rhd}

\newcommand{\lex}{_\mr{lex}}
\newcommand{\mul}{_\mr{mul}}

\renewcommand{\u}[1]{{\underline{#1}}}

\renewcommand{\b}{\beta}
\newcommand{\g}{\gamma}
\newcommand{\G}{\Gamma}
\renewcommand{\d}{\delta}
\newcommand{\D}{\Delta}
\newcommand{\ep}{\epsilon}
\newcommand{\vep}{\varepsilon}
\newcommand{\z}{\zeta}
\renewcommand{\t}{\theta}

\newcommand{\io}{\iota}
\newcommand{\ka}{\kappa}
\newcommand{\la}{\lambda}

\newcommand{\s}{\sigma}

\newcommand{\up}{\upsilon}

\newcommand{\vphi}{\varphi}
\newcommand{\w}{\omega}
\newcommand{\W}{\Omega}

\newcommand{\mi}{\mathit}
\newcommand{\mc}{\mathcal}

\newcommand{\mr}{\mathrm}

\newcommand{\mf}{\mathfrak}

\newcommand{\cA}{\mc{A}}

\newcommand{\cC}{\mc{C}}
\newcommand{\cD}{\mc{D}}

\newcommand{\cF}{\mc{F}}
\newcommand{\cG}{\mc{G}}
\newcommand{\cH}{\mc{H}}

\newcommand{\cK}{\mc{K}}

\newcommand{\cN}{\mc{N}}
\newcommand{\cO}{\mc{O}}
\newcommand{\cP}{\mc{P}}

\newcommand{\cR}{\mc{R}}
\newcommand{\cS}{\mc{S}}
\newcommand{\cT}{\mc{T}}

\newcommand{\cV}{\mc{V}}
\newcommand{\cW}{\mc{W}}
\newcommand{\cX}{\mc{X}}

\newcommand{\cZ}{\mc{Z}}

\newcommand{\fa}{\mf{a}} % amssymb
\newcommand{\fb}{\mf{b}}
\newcommand{\fc}{\mf{c}}

\newcommand{\fA}{\mf{A}}

\newcommand{\va}{{\vec{a}}}

\newcommand{\vl}{{\vec{l}}}

\newcommand{\vs}{{\vec{s}}}
\newcommand{\vt}{{\vec{t}}}
\newcommand{\vu}{{\vec{u}}}
\newcommand{\vv}{{\vec{v}}}
\newcommand{\vw}{{\vec{w}}}
\newcommand{\vx}{{\vec{x}}}
\newcommand{\vy}{{\vec{y}}}
\newcommand{\vz}{{\vec{z}}}

\newcommand{\vQ}{{\vec{Q}}}

\newcommand{\vS}{{\vec{S}}}
\newcommand{\vT}{{\vec{T}}}
\newcommand{\vU}{{\vec{U}}}
\newcommand{\vV}{{\vec{V}}}

\newenvironment{rul}%
  {$\begin{array}{rcl}}%
  {\end{array}$}
  {\begin{center}\begin{rul}}%
  {\end{rul}\end{center}}

\newenvironment{rew}[1][~~\a~~]%
  {$\begin{array}{r@{#1}l}}%
  {\end{array}$}
\newenvironment{rewc}[1][~~\a~~]%
  {\begin{center}\begin{rew}[#1]}%
  {\end{rew}\end{center}}

\newcounter{counter}

\newcounter{explnum}
{\theorembodyfont{\rmfamily} % theorem
  \newtheorem{dfn}[counter]{Definition}
  \newtheorem{lem}[counter]{Lemma}
  \newtheorem{thm}[counter]{Theorem}

  \newtheorem{expl}[explnum]{Example}
}

\newcommand{\cqfd}{\hfill$\blacksquare$} % amssymb
\newenvironment{prf}{{\bf Proof.}}{}

\newenvironment{lstgeneric}[2]
  {\begin{list}{#1}{\topsep=.5mm\itemsep=.5mm\parsep=0mm%
    \itemindent=-3ex\labelsep=1ex\labelwidth=0ex #2}}
  {\end{list}}

\newenvironment{lst}[1]
  {\begin{lstgeneric}{#1}{\itemindent=-1ex}}
  {\end{lstgeneric}}

\newenvironment{enumi}[1]
  {\begin{lstgeneric}{}{\usecounter{enumi}\leftmargin=7mm%
    }}
  {\end{lstgeneric}}

\newenvironment{bfenumii}[1]
  {\begin{lstgeneric}{}{\usecounter{enumii}\leftmargin=7mm%
    }}
  {\end{lstgeneric}}

\newenvironment{bfenumalphai}
  {\begin{lstgeneric}{}{\usecounter{enumi}\leftmargin=7mm%
    }}
  {\end{lstgeneric}}

%% file: macros2.tex
\newcommand{\tf}{{\tau_f}}
\newcommand{\tg}{{\tau_g}}

\newcommand{\tC}{{\tau_C}}

\newcommand{\FVB}{\FV^\B}
\newcommand{\acc}{\mr{Acc}}
\newcommand{\mon}{\mr{Mon}}
\newcommand{\co}{\mr{Cons}}
\newcommand{\Xs}{\cX^\st}
\newcommand{\XB}{\cX^\B}
\newcommand{\domB}{\dom^\B}
\newcommand{\SN}{\cS\cN}
\newcommand{\WN}{\cW\cN}
\newcommand{\DF}{\cD\cF}
\newcommand{\DFB}{\DF^\B}
\newcommand{\CF}{\cC\cF}
\newcommand{\CFB}{\CF^\B}
\newcommand{\Fs}{\cF^\st}
\newcommand{\FB}{\cF^\B}
\newcommand{\CT}{\cC\cT}

\newcommand{\pp}{\pos^+}
\renewcommand{\pm}{\pos^-}
\newcommand{\pd}{\pos^\d}
\newcommand{\pmd}{\pos^{-\d}}

\newcommand{\dgd}[1][]{{\,\tgt\hs[-2.1mm]\cdot\hs[1mm]}}

\newcommand{\nf}{\!\!\ad}

\newcommand{\thc}{\th_\mr{\!\!c}}

\newcommand{\si}{\mi{if}}
\newcommand{\fst}{\mi{fst}}

\newenvironment{trans}
  {\begin{center}$\begin{array}{rr@{~~\a~~}l}}%
  {\end{array}$\end{center}}

%% file: title.tex
%%%%%%%%%%%%%%%%%%%%%%%%%%%%%%%%%%%%%%%%%%%%%%%%%%%%%%%%%%%%%%%%%%%%%%%%%%%%%%
% title
%%%%%%%%%%%%%%%%%%%%%%%%%%%%%%%%%%%%%%%%%%%%%%%%%%%%%%%%%%%%%%%%%%%%%%%%%%%%%%

\title{A type-based termination criterion for dependently-typed higher-order rewrite systems}

\author{Fr\'ed\'eric Blanqui \footnote{
Laboratoire Lorrain de Recherche en Informatique et Automatique (LORIA)
\& Institut National de Recherche en Informatique et Automatique (INRIA),
615 rue du Jardin Botanique, BP 101, 54602 Villers-l\`es-Nancy, France,
{\tt blanqui@loria.fr}.}}

\date{January 22, 2004}

\maketitle

\noindent{\bf Abstract:} {\it Several authors devised
type-based termination criteria for ML-like languages (polymorphic
$\la$-calculi with inductive types and case analysis), that allows
non-structural recursive calls. We extend these works to general
rewriting and dependent types, hence providing a powerful termination
criterion for the combination of rewriting and $\b$-reduction in the
Calculus of Constructions.}

%% file: intro.tex
%%%%%%%%%%%%%%%%%%%%%%%%%%%%%%%%%%%%%%%%%%%%%%%%%%%%%%%%%%%%%%%%%%%%%%%%%%%%%%
% introduction
%%%%%%%%%%%%%%%%%%%%%%%%%%%%%%%%%%%%%%%%%%%%%%%%%%%%%%%%%%%%%%%%%%%%%%%%%%%%%%

\section{Introduction}

The Calculus of Constructions \cite{coquand88ic} is a powerful type
system allowing polymorphic and dependent types. It is the basis of
many proof assistants since it allows one to formalize the proofs of
higher-order logic. In this context, it is essential to allow users to
define functions and predicates in the most convenient way and to be
able to decide whether a term is a proof of some proposition, and
whether two terms/propositions are equivalent w.r.t. user
definitions. As exemplified in \cite{dowek98types,blanqui03mscs}, a
promising approach is rewriting. To this end, we need powerful
criteria to check the termination of higher-order rewrite-based
definitions combined with $\b$-reduction.

In a previous work \cite{blanqui03mscs}, we proved that such a
combination is strongly normalizing if, on the one hand, first-order
rewrite rules are strongly normalizing and
non-duplicating\footnote{Strong normalization is not modular in
general \cite{toyama87ipl}. It is modular for non-duplicating
first-order rewrite systems \cite{rusinowitch87ipl}. Here, we do not
have two non-duplicating first-order rewrite systems but a
hierarchical combination of a higher-order rewrite system (satisfying
strong termination conditions) built over a non-duplicating
first-order rewrite system.} and, on the other hand, non first-order
rewrite rules (called higher-order in the following) satisfies a
termination criterion based on the notion of computability closure and
similar to higher-order primitive recursion.  Unfortunately, many
interesting rewrite systems are either first-order and duplicating, or
higher-order with non-structural recursive calls ({\em e.g.} division
on natural numbers\footnote{$/~x~y$ denotes
$\lceil\frac{x}{y+1}\rceil$.}\footnote{We use curried symbols all over
the paper.}, Figure \ref{fig-div}).

\begin{figure}[ht]
\caption{Division on natural numbers\label{fig-div}}
\begin{trans}
(1) & -~x~0 & x\\
(2) & -~0~x & 0\\
(3) & -~(sx)~(sy) & -~x~y\\[2mm]

(4) & /~0~x & 0\\
(5) & /~(sx)~y & s~(/~(-~x~y)~y)\\
\end{trans}
\end{figure}

Hughes {\em et al} \cite{hughes96popl}, Xi \cite{xi01lics,xi02hosc},
Gim\'enez {\em et al} \cite{gimenez98icalp,barthe04mscs} and Abel
\cite{abel02tr} devised termination criteria able to treat such
examples by exploiting the way inductive types are usually interpreted
\cite{mendler87thesis}. Take for instance the
addition\footnote{$[x:T]u$ denotes the function which associates $u$
to every $x$ of type $T$.} on Brouwer's ordinals $ord$ (Figure
\ref{fig-ord}) whose constructors are $0:ord$, $s:ord\A ord$ and
$lim:(nat\A ord)\A ord$.

\begin{figure}[ht]
\caption{Addition on Brouwer's ordinals\label{fig-ord}}
\begin{trans}
(1) & +~0~x & x\\
(2) & +~(s x)~y & s~(+~x~y)\\
(3) & +~(lim~f)~y & lim~([x:nat](+~(f~x)~y))\\
\end{trans}
\end{figure}

The usual computability-based technique for proving the termination of
this function is to interpret $ord$ by the fixpoint of the following
monotone function $\vphi$ on the powerset of $\SN$, the set of
strongly normalizing terms, ordered by inclusion:\footnote{$\a^*$ is
the reflexive and transitive closure of the reduction relation $\a$.}

\begin{center}
$\vphi(X)= \{t\in\SN~|~ t\a^* su\A u\in X; t\a^* lim f\A \all
u\in\SN, fu\in X\}$
\end{center}

The fixpoint of $\vphi$, $\I{ord}$, can be reached by transfinite
iteration and every $t\in\I{ord}$ is obtained after a smallest ordinal
$o(t)$ of iterations, the order of $t$. This naturally defines an
ordering: $t>u$ iff $o(t)>o(u)$, with which we clearly have $lim~f>fu$
for all $u\in\SN$.

Now, applying this technique to $nat$, we can easily check that
$o(-tu)\le o(t)$ and thus allow the recursive call with $-xy$ in the
definition of $/$. First note that $-tu$ is computable ({\em i.e.}
belongs to $\I{nat}$) iff all its reducts are computable (see Section
\ref{sec-sn}). We proceed by induction on $o(t)$:

\begin{lst}{--}
\item If $-tu$ matches rule (1) then $o(-tu)=o(t)$.
\item If $-tu$ matches rule (2) then $o(-tu)=0\le o(t)$.
\item If $-tu$ matches rule (3) then $t=st'$ and $u=su'$. By induction
  hypothesis, $o(-t'u')\le o(t')$. Thus, $o(-tu)=1+o(-t'u')\le
  1+o(t')=o(t)$.
\item If $-tu$ matches no rule then $o(-tu)=0\le o(t)$.
\end{lst}

The idea of the previously cited authors is to add this
size/index/stage information to the syntax in order to prove this
automatically. Instead of a single type $nat$, they consider a family
of types $\{nat^\fa\}_{\fa\in\w}$, each type $nat^\fa$ being
interpreted by the set obtained after $\fa$ iterations of the function
$\vphi$ for $nat$.  And they define a decidable type system in which
minus (defined by {\em fixpoint/cases} constructions in their work)
can be typed by $nat^\alpha\A nat^\beta\A nat^\alpha$, where $\alpha$
and $\beta$ are size variables, meaning that the order of $-tu$ is not
greater than the order of $t$.

This can also be interpreted as a way to automatically prove theorems
on the size of the result of a function w.r.t. the size of its
arguments \cite{walther88cade,giesl97jar} with application to
complexity and resource bound certification, and compilation
optimization ({\em e.g.} bound check elimination
\cite{pfenning98pldi}, vector-based memoisation \cite{chin97icfp}).

In this paper, we extend this technique to the full Calculus of
Algebraic Constructions \cite{blanqui03mscs} whose type conversion
rule depends on the user-defined rewrite rules, and to general
rewrite-based definitions (including matching on defined symbols and
rewriting modulo equational theories \cite{blanqui03rta}) instead of
definitions only based on $letrec/match$ (or $fixpoint/cases$)
constructions. Note that our work makes a heavy use of (and simplify)
the techniques developed by Chen for studying the Calculus of
Constructions with subtyping \cite{chen98thesis}.

On the one hand, we allow a richer size algebra than the one in
\cite{hughes96popl,barthe04mscs,abel02tr} (see Section \ref{sec-cons}). On
the other hand, we do not allow existential size variables and
conditional rewriting\footnote{The equivalent of {\em if-then-else}
constructions in functional programming.} that are essential for
capturing, for instance, the size-preserving property of quicksort
(Example \ref{expl-qs}) and Mac Carty's ``91'' function (Example
\ref{expl-mc}) respectively, as it can be done in Xi's work
\cite{xi02hosc}. Note however that Xi is interested in the call-by-value
normalization of closed simply-typed $\la$-terms, while we are
interested in the strong normalization of the open terms of the
Calculus of Constructions.

%% file: cac.tex
%%%%%%%%%%%%%%%%%%%%%%%%%%%%%%%%%%%%%%%%%%%%%%%%%%%%%%%%%%%%%%%%%%%%%%%%%%%%%%
% CACSA
%%%%%%%%%%%%%%%%%%%%%%%%%%%%%%%%%%%%%%%%%%%%%%%%%%%%%%%%%%%%%%%%%%%%%%%%%%%%%%

\section{The Calculus of Algebraic Constructions with Size Annotations}
\label{sec-cacsa}

The Calculus of Constructions (CC) is the full Pure Type System with the
set of {\em sorts} $\cS=\{\st,\B\}$ and the axiom $\st:\B$
\cite{barendregt92book}. $\st$ is intended to be the universe of types and
propositions, while $\B$ is intended to be the universe of predicate
types. Let $\cX$ be the set of variables.

The Calculus of Algebraic Constructions (CAC) \cite{blanqui03mscs} is
an extension of CC with a set $\cF$ of function or predicate {\em
symbols} defined by a set $\cR$ of (higher-order) rewrite rules
\cite{dershowitz90book,klop93tcs}. Every variable $x$ (resp. symbol
$f$) is equipped with a sort $s_x$ (resp. $s_f$). We denote by $\DF$
the set of {\em defined} symbols, that is, the set of symbols $f$ such
that there is a rule $l\a r\in\cR$ with $l=f\vl$, and by $\CF$ the set
$\cF\moins\DF$ of {\em constant} symbols. We add a superscript $s$ to
restrict these sets to variables or symbols of sort $s$.

%%%%%%%%%%%%%%%%%%%%%%%%%%%%%%%%%%%%%%%%%%%%%%%%%%%%%%%%%%%%%%%%%%%%%%%%%%%%%%
% size expressions
%%%%%%%%%%%%%%%%%%%%%%%%%%%%%%%%%%%%%%%%%%%%%%%%%%%%%%%%%%%%%%%%%%%%%%%%%%%%%%

Now, we assume given a (sorted) first-order term algebra
$\cA=T(\cH,\cZ)$, called the algebra of {\em size expressions}, built
from a non-empty set $\cH$ of {\em size symbols} of fixed arity and a
set $\cZ$ of {\em size variables}. We assume that $\cH\cap\cF=
\cZ\cap\cX= \vide$. Let $\cV(t)$ be the set of size variables
occurring in a term $t$. A {\em renaming} is an injection from a
finite subset of $\cZ$ to $\cZ$.

We assume that, for every rule $l\a r\in\cR$, $\cV(l)= \cV(r)=
\vide$. Hence, if $t\a t'$ then, for all size substitution $\vphi$,
$t\vphi\a t'\vphi$.

We also assume that $\cA$ is equipped with a quasi-ordering $\le_\cA$
stable by size substitution ({\em i.e.} if $a\le_\cA b$ then, for all
size substitution $\vphi$, $a\vphi\le_\cA b\vphi$) such that
$(\cA,\le_\cA)$ has a well-founded model $(\fA,\le_\fA)$:

\begin{dfn}[Size model]
A {\em pre-model} of $\cA$ is given by a set $\fA$, an ordering
$\le_\fA$ on $\fA$ and a function $h_\fA$ from $\fA^n$ to $\fA$ for
every $n$-ary size symbol $h\in\cH$. A {\em size valuation} is a
function $\nu$ from $\cZ$ to $\fA$, naturally extended to a function
on $\cA$. A pre-model is a {\em model} if, for all size valuation
$\nu$, $a\nu\le_\fA b\nu$ whenever $a\le_\cA b$. Such a model is {\em
well-founded} if $>_\fA$ is well-founded.
\end{dfn}

%%%%%%%%%%%%%%%%%%%%%%%%%%%%%%%%%%%%%%%%%%%%%%%%%%%%%%%%%%%%%%%%%%%%%%%%%%%%%%
% terms
%%%%%%%%%%%%%%%%%%%%%%%%%%%%%%%%%%%%%%%%%%%%%%%%%%%%%%%%%%%%%%%%%%%%%%%%%%%%%%

The Calculus of Algebraic Constructions with Size Annotations (CACSA)
is an extension of CAC where constant predicate symbols are annotated
by size expressions. The terms of CACSA are defined by the following
grammar rule:

\begin{center}
$t ::= s ~|~ x ~|~ C^a ~|~ f ~|~ [x:t]t ~|~ (x:t)t ~|~ tt$
\end{center}

\noindent
where $C\in\CFB$, $f\in\cF\moins\CFB$ and $a\in\cA$. We denote by
$\cT_\cA(\cF,\cX)$ the set of terms built from $\cF$, $\cX$ and
$\cA$. Let $\u\cT$ be the set of the underlying CAC terms and $\u~$ be
the function erasing size annotations. Among CAC terms, we distinguish
the following disjoint sets:

\begin{lst}{--}
\item {\em kinds}: $K\in\cK ::= \st ~|~ (x:t)K$
\item {\em predicates}: $P\in\cP ::= f\in\FB ~|~ x\in\XB ~|~ (x:t)P
~|~ [x:t]P ~|~ Pt$
\item {\em objects}: $o\in\cO ::= f\in\Fs ~|~ x\in\Xs ~|~ [x:t]o ~|~ ot$
\end{lst}

\noindent
where $t\in\u\cT$ is any CAC term.

%%%%%%%%%%%%%%%%%%%%%%%%%%%%%%%%%%%%%%%%%%%%%%%%%%%%%%%%%%%%%%%%%%%%%%%%%%%%%%
% typing
%%%%%%%%%%%%%%%%%%%%%%%%%%%%%%%%%%%%%%%%%%%%%%%%%%%%%%%%%%%%%%%%%%%%%%%%%%%%%%

Finally, we assume that every symbol $f$ is equipped with a type
$\tf={(\vx:\vT)}U\in\cT$ such that $\FV(\tf)=\vide$, $s_f=\B\A
\cV(\tf)=\vide$, and $f\vl\a r\in\cR\A |\vl|\le|\vt|$.

We also assume that every symbol $f$ is equipped with a set
$\mon^+(f)\sle A_f= \{1,\ldots,|\vx|\}$ of {\em monotone arguments}
and a set $\mon^-(f)\sle A_f$ of {\em anti-monotone arguments} such
that $\mon^+(f)\cap\mon^-(f)=\vide$. For a size symbol $h$,
$\mon^+(h)$ (resp. $\mon^-(h)$) is taken to be the arguments in which
$h_\fA$ is monotone (resp. anti-monotone).

An {\em environment} $\G$ is a sequence of pairs variable-term. Let
$t\ad u$ iff there is $v$ such that $t\a^* v\als u$. The typing rules
of CACSA are given in Figure \ref{fig-typ} and its subtyping rules in
Figure \ref{fig-sub}. W.l.o.g. we can assume that, for all $f$,
$\th\tf:s_f$. We also assume that, for every rule $l\a r\in\cR$, there
exist an environment $\G$ and a type $T$ such that $\G\th r:T$. This
is to make sure that $r$ is not ill-formed (see Lemma 12 in
\cite{blanqui03mscs}).

Since, in the (symb) rule, symbol types are applied to arbitrary size
substitutions $\vphi$, the name of size variables in symbol types is
not relevant (size variables in symbol types are implicitly
universally quantified).

A substitution $\t$ {\em preserves typing} between $\G$ and $\D$,
written $\t:\G\leadsto\D$, iff $\D\th x\t:x\G\t$ for all
$x\in\dom(\G)$. A type-preserving substitution satisfies the following
important substitution property: if $\G\th t:T$ and $\t:\G\leadsto\D$
then $\D\th t\t:T\t$.

%%%%%%%%%%%%%%%%%%%%%%%%%%%%%%%%%%%%%%%%%%%%%%%%%%%%%%%%%%%%%%%%%%%%%%%%%%%%%%
% subtyping rules
%%%%%%%%%%%%%%%%%%%%%%%%%%%%%%%%%%%%%%%%%%%%%%%%%%%%%%%%%%%%%%%%%%%%%%%%%%%%%%

\begin{figure}[ht]
\centering
\caption{Subtyping rules\label{fig-sub}}
\begin{tabular}{rccl}

\\ (refl) & $T\le T$\\

\\ (size) & $C^a\vt\le C^b\vt$ & ($C\in\CFB$, $a\le_\cA b$)\\

\\ (prod) & $\cfrac{U'\le U \quad V\le V'}{(x:U)V\le (x:U')V'}$\\

\\ (conv) & $\cfrac{T'\le U'}{T\le U}$
& ($T\ad T'$, $U'\ad U$)\\

\\ (trans) & $\cfrac{T\le U\quad U\le V}{T\le V}$\\
\end{tabular}
\end{figure}

%%%%%%%%%%%%%%%%%%%%%%%%%%%%%%%%%%%%%%%%%%%%%%%%%%%%%%%%%%%%%%%%%%%%%%%%%%%%%%
% typing rules
%%%%%%%%%%%%%%%%%%%%%%%%%%%%%%%%%%%%%%%%%%%%%%%%%%%%%%%%%%%%%%%%%%%%%%%%%%%%%%

\begin{figure}[ht]
\centering
\caption{Typing rules\label{fig-typ}}
\begin{tabular}{rcc}
\\ (ax) & $\th\st:\B$\\

\\ (size) & $\cfrac{\th\tC:\B}{\th C^a:\tC}$ & ($C\in\CFB$)\\

\\ (symb) & $\cfrac{\th\tf:s_f}{\th f:\tf\vphi}$ & ($f\notin\CFB$)\\

\\ (var) & $\cfrac{\G\th T:s_x}{\G,x:T\th x:T}$
& $(x\notin\dom(\G))$\\

\\ (weak) & $\cfrac{\G\th t:T \quad \G\th U:s_x}{\G,x:U\th t:T}$
& $(x\notin\dom(\G))$\\

\\ (prod) & $\cfrac{\G\th U:s \quad \G,x:U \th V:s'}
{\G\th (x:U)V:s'}$\\

\\ (abs) & $\cfrac{\G,x:U \th v:V \quad \G\th (x:U)V:s}
{\G\th [x:U]v:(x:U)V}$\\

\\ (app) & $\cfrac{\G\th t:(x:U)V \quad \G\th u:U}
{\G\th tu:V\xu}$\\

\\ (sub) & $\cfrac{\G\th t:T \quad \G\th T':s}{\G\th t:T'}$ & ($T\le T'$)\\
\end{tabular}
\end{figure}

%%%%%%%%%%%%%%%%%%%%%%%%%%%%%%%%%%%%%%%%%%%%%%%%%%%%%%%%%%%%%%%%%%%%%%%%%%%%%%
% assumptions
%%%%%%%%%%%%%%%%%%%%%%%%%%%%%%%%%%%%%%%%%%%%%%%%%%%%%%%%%%%%%%%%%%%%%%%%%%%%%%

In this paper, we make two important assumptions.\\

\noindent{\bf Assumptions:}
\begin{enumi}{}
\item $\b\cup\cR$ is confluent. This is the case for instance if $\cR$
is confluent and left-linear. Finding other sufficient conditions when
there are type-level rewrite rules is an open problem.

\item $\cR$ preserves typing: if $l\a r\in\cR$ and $\G\th l\s:T$ then
$\G\th r\s:T$. Finding sufficient conditions with subtyping and
dependent types does not seem easy as shown by the following
example. We leave the study of this problem for future work.
\end{enumi}

%%%%%%%%%%%%%%%%%%%%%%%%%%%%%%%%%%%%%%%%%%%%%%%%%%%%%%%%%%%%%%%%%%%%%%%%%%%%%%
% subject reduction
%%%%%%%%%%%%%%%%%%%%%%%%%%%%%%%%%%%%%%%%%%%%%%%%%%%%%%%%%%%%%%%%%%%%%%%%%%%%%%

\begin{expl}[Subject reduction]
\label{expl-sr}
Assume that $s\in\cH$, $nat:\st$, $s:nat^\alpha\A nat^{s\alpha}$,
$-:nat^\alpha\A nat^\beta\A nat^\alpha$, and let us prove that the
rule $-(sx)(sy)\a -xy$ preserves typing. Assume that $\G\th
-(st)(su):T$. We must prove that $\G\th -t u:T$. By inversion,
$\G\th -(st):(z_2:T_2)U_2$, $\G\th su:T_2$ and $U_2\{z_2\to
su\}\le T$. By inversion again, $\G\th -:(z_1:T_1)U_1$, $\G\th
st:T_1$ and $U_1\{z_1\to st\}\le (z_2:T_2)U_2$. Again, $nat^a\A
nat^b\A nat^a\le (z_1:T_1)U_1$, $\G\th s:(z_3:T_3)U_3$, $\G\th
t:T_3$, $U_3\{z_3\to t\}\le T_1$, $nat^c\A nat^{sc}\le
(z_3:T_3)U_3$, $\G\th s:(z_4:T_4)U_4$, $\G\th u:T_4$, $U_4\{z_4\to
u\}\le T_2$ and $nat^d\A nat^{sd}\le (z_4:T_4)U_4$. By Lemma
\ref{lem-prod-comp}, we have $T_3\le nat^c$, $nat^{sc}\le U_3$,
$T_4\le nat^d$, $nat^{sd}\le U_4$, $T_1\le nat^a$ and $nat^b\A
nat^a\le U_1$. Again, since $U_1\{z_1\to st\}\le (z_2:T_2)U_2$,
$T_2\le nat^b$ and $nat^a\le U_2$. Therefore, since $\G\th t:T_3\le
nat^c$, $\G\th u:T_4\le nat^d$ and $\G\th -:nat^c\A nat^d\A nat^c$, we
have $\G\th -t u:nat^c$. Now, we must prove that $nat^c\le T$. First,
$nat^c\le nat^{sc}\le U_3$. Since $U_3\{z_3\to t\}\le T_1$, $nat^c\le
T_1$. Since $nat^a\A nat^b\A nat^a\le (z_1:T_1)U_1$, $T_1\le nat^a$
and $nat^b\A nat^a\le U_1$. Since $U_1\{z_1\to st\}\le (z_2:T_2)U_2$,
$nat^b\A nat^a\le (z_2:T_2)U_2$. Therefore, $nat^a\le U_2$. Now, since
$U_2\{z_2\to su\}\le T$, we indeed have $nat^c\le T$.
\end{expl}

%% file: sub.tex
%%%%%%%%%%%%%%%%%%%%%%%%%%%%%%%%%%%%%%%%%%%%%%%%%%%%%%%%%%%%%%%%%%%%%%%%%%%%%%
% properties of subtyping
%%%%%%%%%%%%%%%%%%%%%%%%%%%%%%%%%%%%%%%%%%%%%%%%%%%%%%%%%%%%%%%%%%%%%%%%%%%%%%

\section{Properties of subtyping}

%%%%%%%%%%%%%%%%%%%%%%%%%%%%%%%%%%%%%%%%%%%%%%%%%%%%%%%%%%%%%%%%%%%%%%%%%%%%%%
% size substitution
%%%%%%%%%%%%%%%%%%%%%%%%%%%%%%%%%%%%%%%%%%%%%%%%%%%%%%%%%%%%%%%%%%%%%%%%%%%%%%

\begin{lem}
If $U\le V$ then, for all size substitution $\psi$, $U\psi\le V\psi$.
\end{lem}

\begin{prf}
Easy induction.\cqfd\\
\end{prf}

%%%%%%%%%%%%%%%%%%%%%%%%%%%%%%%%%%%%%%%%%%%%%%%%%%%%%%%%%%%%%%%%%%%%%%%%%%%%%%
% elimination of transitivity
%%%%%%%%%%%%%%%%%%%%%%%%%%%%%%%%%%%%%%%%%%%%%%%%%%%%%%%%%%%%%%%%%%%%%%%%%%%%%%

We now prove that the subtyping rule (trans) can be eliminated.

\begin{thm}[Transitivity elimination]
\label{thm-trans-elim}
Let $\le_t$ be the subtyping relation obtained without using
(trans). Then, $\le_t=\le$.
\end{thm}

\begin{prf}
Section \ref{sec-trans}.\cqfd\\
\end{prf}

This means that, in a subtyping derivation, we can always assume that
there is no application of (trans) and that, in a typing derivation,
there is no successive applications of (sub).

%%%%%%%%%%%%%%%%%%%%%%%%%%%%%%%%%%%%%%%%%%%%%%%%%%%%%%%%%%%%%%%%%%%%%%%%%%%%%%
% product compatibility
%%%%%%%%%%%%%%%%%%%%%%%%%%%%%%%%%%%%%%%%%%%%%%%%%%%%%%%%%%%%%%%%%%%%%%%%%%%%%%

\begin{lem}[Product compatibility]
\label{lem-prod-comp}
If $(x:U)V\le (x:U')V'$ then $U'\le U$ and $V\le V'$.
\end{lem}

\begin{prf}
By case on the last rule of $(x:U)V\le (x:U')V'$. By confluence, we
can assume that there is no successive applications of (conv). This is
immediate for (refl) and (prod). (symb) is not possible. For (conv),
we have:

\begin{center}
$\cfrac{(x:U)V\ad T\le T'\ad (x:U')V'}{(x:U)V\le (x:U')V'}$
\end{center}

Then, we reason by case on the last rule of $T\le T'$.

\begin{lst}{}
\item [\bf(refl)] In this case, $T=T'$. Therefore,
by confluence, $(x:U)V\ad (x:U')V'$, $U\ad U'$ and $V\ad V'$. Thus,
$U'\le U$ and $V\le V'$.

\item [\bf(symb)] Not possible since $T=C^a\vt$ has
no common reduct with $(x:U)V$ (since $C$ is constant).

\item [\bf(conv)] Excluded.

\item [\bf(prod)] In this case, $T=(x:U_1)V_1$,
$T'=(x:U_2)V_2$, $U_2\le U_1$ and $V_1\le V_2$. By confluence $U\ad
U_1$, $V\ad V_1$, $U_2\ad U'$ and $V_2\ad V'$. Therefore, by
conversion, $U'\le U$ and $V\le V'$.\cqfd\\
\end{lst}
\end{prf}

%%%%%%%%%%%%%%%%%%%%%%%%%%%%%%%%%%%%%%%%%%%%%%%%%%%%%%%%%%%%%%%%%%%%%%%%%%%%%%
% expansion elimination
%%%%%%%%%%%%%%%%%%%%%%%%%%%%%%%%%%%%%%%%%%%%%%%%%%%%%%%%%%%%%%%%%%%%%%%%%%%%%%

We now prove that the subtyping relation can be further
simplified. Consider the following two admissible rules:

\begin{center}
\begin{tabular}{rcc}
(red) & $\cfrac{T\a^* T'\quad T'\le U'\quad U'\als U}{T\le U}$\\
\\ (exp) & $\cfrac{T\als T'\quad T'\le U'\quad U'\a^* U}{T\le U}$\\
\end{tabular}
\end{center}

(conv) can clearly be replaced by both (red) and (exp).

\begin{thm}[Expansion elimination]
\label{thm-exp-elim}
Let $\le_r$ be the subtyping relation with (red) instead of
(conv). Then, $\le_r=\le$.
\end{thm}

\begin{prf}
Section \ref{sec-exp}.\cqfd\\
\end{prf}

%%%%%%%%%%%%%%%%%%%%%%%%%%%%%%%%%%%%%%%%%%%%%%%%%%%%%%%%%%%%%%%%%%%%%%%%%%%%%%
% subtyping without conversion
%%%%%%%%%%%%%%%%%%%%%%%%%%%%%%%%%%%%%%%%%%%%%%%%%%%%%%%%%%%%%%%%%%%%%%%%%%%%%%

Now, let $\le_s$ be the subtyping relation with (refl), (symb) and
(prod) only.

\begin{lem}
\label{lem-sub-s}
$T\le U$ iff there exist $T'$ and $U'$ such that $T\a^* T'\le_s U'\als
U$. Furthermore, if $T,U\in\WN$ then $T\nf\le_s U\nf$.
\end{lem}

\begin{prf}
The if-part is immediate. The only-if-part is easily proved by
induction on $T\le U$. In the (red) case, if $T\a^* T'\le U'\als U$
then, by induction hypothesis, there exist $T''$ and $U''$ such that
$T'\a^* T''\le_s U''\als U'$. Therefore, $T\a^* T''\le_s U''\als U$.

Now, if $T,U\in\WN$ then $T\nf\le U\nf$. Thus, $T\nf\le_s U\nf$ since
$T\nf$ and $U\nf$ are not reducible.\cqfd
\end{prf}

%%%%%%%%%%%%%%%%%%%%%%%%%%%%%%%%%%%%%%%%%%%%%%%%%%%%%%%%%%%%%%%%%%%%%%%%%%%%%%
% subtyping kinds
%%%%%%%%%%%%%%%%%%%%%%%%%%%%%%%%%%%%%%%%%%%%%%%%%%%%%%%%%%%%%%%%%%%%%%%%%%%%%%

\begin{lem}
\label{lem-sub-kinds}
\begin{lst}{--}
\item For all $s\in\cS$, if $T\le s$ or $s\le T$ then $T\a^* s$.
\item For all $K\in\cK$, if $T\le K$ or $K\le T$ then $T\a^* T'\in\cK$.
\end{lst}
\end{lem}

\begin{prf}
\begin{lst}{--}
\item If $s\le T$ then $s\le_s T'\als T$. The only possible case
is $T'=s$. If $T\le s$ then $T\a^* T'\le_s s$. The only possible case
is $T'=s$.
\item If $T\le K$ then $T\a^* T'\le_s K'\als K$ and $K'\in\cK$.
Now, one can easily prove by induction that, if $T'\le_s K'$, then
$T'\in\cK$. If $K\le T$ then $K\a^* K'\le_s T'\als T$ and $K'\in\cK$.
One can easily prove by induction that, if $K'\le_s T'$, then
$T'\in\cK$.\cqfd
\end{lst}
\end{prf}

%%%%%%%%%%%%%%%%%%%%%%%%%%%%%%%%%%%%%%%%%%%%%%%%%%%%%%%%%%%%%%%%%%%%%%%%%%%%%%
% decidability
%%%%%%%%%%%%%%%%%%%%%%%%%%%%%%%%%%%%%%%%%%%%%%%%%%%%%%%%%%%%%%%%%%%%%%%%%%%%%%

\begin{thm}[Decidability of subtyping]
$\le$ is decidable whenever $\a$ is confluent, weakly normalizing and
finitely branching (or confluent and strongly normalizing).
\end{thm}

\begin{prf}
Immediate consequence of Lemma \ref{lem-sub-s}.
\end{prf}

%% file: typ.tex
%%%%%%%%%%%%%%%%%%%%%%%%%%%%%%%%%%%%%%%%%%%%%%%%%%%%%%%%%%%%%%%%%%%%%%%%%%%%%%
% properties of typing
%%%%%%%%%%%%%%%%%%%%%%%%%%%%%%%%%%%%%%%%%%%%%%%%%%%%%%%%%%%%%%%%%%%%%%%%%%%%%%

\section{Properties of typing}

%%%%%%%%%%%%%%%%%%%%%%%%%%%%%%%%%%%%%%%%%%%%%%%%%%%%%%%%%%%%%%%%%%%%%%%%%%%%%%
% size substitution
%%%%%%%%%%%%%%%%%%%%%%%%%%%%%%%%%%%%%%%%%%%%%%%%%%%%%%%%%%%%%%%%%%%%%%%%%%%%%%

\begin{lem}
\label{lem-size-sub}
If $\G\th t:T$ then, for all size substitution $\psi$, $\G\psi\th
t\psi:T\psi$.
\end{lem}

\begin{prf}
Easy induction.\cqfd
\end{prf}

%%%%%%%%%%%%%%%%%%%%%%%%%%%%%%%%%%%%%%%%%%%%%%%%%%%%%%%%%%%%%%%%%%%%%%%%%%%%%%
% type correctness
%%%%%%%%%%%%%%%%%%%%%%%%%%%%%%%%%%%%%%%%%%%%%%%%%%%%%%%%%%%%%%%%%%%%%%%%%%%%%%

\begin{lem}[Type correctness]
\label{lem-typ-cor}
If $\G\th t:T$ then either $T=\B$ or $\G\th T:s$ for some sort $s$.
\end{lem}

\begin{prf}
Easy induction.\cqfd
\end{prf}

%%%%%%%%%%%%%%%%%%%%%%%%%%%%%%%%%%%%%%%%%%%%%%%%%%%%%%%%%%%%%%%%%%%%%%%%%%%%%%
% properties on kinds, etc.
%%%%%%%%%%%%%%%%%%%%%%%%%%%%%%%%%%%%%%%%%%%%%%%%%%%%%%%%%%%%%%%%%%%%%%%%%%%%%%

\begin{lem}
\label{lem-kinds}
\begin{lst}{--}
\item If $T\a^*\B$ then $T$ is not typable.
\item If $\G\th t:\B$ then $t\in\cK$.
\item If $K\in\cK$ and $\G\th K:L$ then $L=\B$.
\item If $T\a^* K\in\cK$ and $\G\th T:s$ then $T\in\cK$ and $s=\B$.
\end{lst}
\end{lem}

\begin{prf}
These properties are proved for CAC in \cite{blanqui03mscs} (Lemma
11). Their proofs need only a few corrections based on Lemma
\ref{lem-sub-kinds} to be valid for CACSA too.\cqfd
\end{prf}

%%%%%%%%%%%%%%%%%%%%%%%%%%%%%%%%%%%%%%%%%%%%%%%%%%%%%%%%%%%%%%%%%%%%%%%%%%%%%%
% narrowing
%%%%%%%%%%%%%%%%%%%%%%%%%%%%%%%%%%%%%%%%%%%%%%%%%%%%%%%%%%%%%%%%%%%%%%%%%%%%%%

\begin{lem}[Narrowing]
If $\G,y:A,\G'\th t:T$, $B\le A$, $\G\th B:s_y$ then $\G,y:B,\G'\th t:T$.
\end{lem}

\begin{prf}
By induction on $\G,y:A,\G'\th t:T$. We only detail some cases.

\begin{lst}{}
\item [\bf(var)] There are two cases. Assume that we have
$\G\th A:s_y$ and $\G,y:A\th y:A$. Since $\G\th B:s_y$, by (var),
$\G,y:B\th y:B$. Since $B\le A$ and $\G\th A:s_y$, by (sub),
$\G,y:B\th y:A$.

Assume now that we have $\G,y:A,\G'\th T:s_x$ and $\G,y:A,\G',x:T\th
x:T$. By induction hypothesis, $\G,y:B,\G'\th T:s_x$. Thus, by (var),
$\G,y:B,\G',x:T\th x:T$.

\item [\bf(weak)] There are two cases. Assume that we have
$\G\th t:T$, $\G\th A:s_y$ and $\G,y:A\th t:T$. Since $\G\th B:s_y$,
by (weak), $\G,y:B\th t:T$.

Assume now that we have $\G,y:A,\G'\th t:T$, $\G,y:A,\G'\th U:s_x$ and
$\G,y:A,\G',x:U\th t:T$. By induction hypothesis, $\G,y:B,\G'\th t:T$
and $\G,y:B,\G'\th U:s_x$. Thus, by (weak), $\G,y:B,\G',x:U\th
t:T$.\cqfd
\end{lst}
\end{prf}

%%%%%%%%%%%%%%%%%%%%%%%%%%%%%%%%%%%%%%%%%%%%%%%%%%%%%%%%%%%%%%%%%%%%%%%%%%%%%%
% subject reduction for beta
%%%%%%%%%%%%%%%%%%%%%%%%%%%%%%%%%%%%%%%%%%%%%%%%%%%%%%%%%%%%%%%%%%%%%%%%%%%%%%

\begin{thm}[$\b$-Subject reduction]
If $\G\th t:T$ and $t\ab t'$ then $\G\th t':T$.
\end{thm}

\begin{prf}
By induction on $\G\th t:T$, we also prove that, if $\G\ab\G'$, then
$\G'\th t:T$. We only detail the case of a $\b$-head reduction. Assume
that we have $\G\th [x:U']v:(x:U)V$ and $\G\th u:U$. We must prove
that $\G\th v\xu:V\xu$. By inversion, $\G,x:U'\th v:V'$, $\G\th
(x:U')V':s'$, $(x:U')V'\le (x:U)V$ and $\G\th (x:U)V:s$. By product
compatibility, $U\le U'$ and $V'\le V$. By inversion, $\G\th U:s_1$
and $\G\th V':s_2$. By narrowing and subtyping, $\G,x:U\th
v:V$. Therefore, by substitution, $\G\th v\xu:V\xu$.\cqfd
\end{prf}

%%%%%%%%%%%%%%%%%%%%%%%%%%%%%%%%%%%%%%%%%%%%%%%%%%%%%%%%%%%%%%%%%%%%%%%%%%%%%%
% conversion and sorting
%%%%%%%%%%%%%%%%%%%%%%%%%%%%%%%%%%%%%%%%%%%%%%%%%%%%%%%%%%%%%%%%%%%%%%%%%%%%%%

\begin{lem}
If $\G\th t:T$, $T\le T'$ and $\G\th T':s'$ then $\G\th T:s$ for some
$s$.
\end{lem}

\begin{prf}
By type correctness, either $T=\B$ or $\G\th T:s$ for some $s$. If
$T=\B$ then, by Lemma \ref{lem-sub-kinds}, $T'\a^*\B$ and, by Lemma
\ref{lem-kinds}, $T'$ cannot be typable.\cqfd
\end{prf}

%%%%%%%%%%%%%%%%%%%%%%%%%%%%%%%%%%%%%%%%%%%%%%%%%%%%%%%%%%%%%%%%%%%%%%%%%%%%%%
% unicity of sorting
%%%%%%%%%%%%%%%%%%%%%%%%%%%%%%%%%%%%%%%%%%%%%%%%%%%%%%%%%%%%%%%%%%%%%%%%%%%%%%

\begin{lem}[Unicity of sorting]
If $T\le T'$, $\G\th T:s$ and $\G\th T':s'$ then $s=s'$.
\end{lem}

\begin{prf}
If $s=\B$ then $T\in\cK$. By Lemma \ref{lem-sub-kinds}, $T'\a^*
K\in\cK$. By Lemma \ref{lem-kinds}, $T'\in\cK$ and $s'=\B$. By
symmetry, if $s'=\B$ then $s=\B$. So, $s=\B$ iff $s'=\B$. Since
$s,s'\in\cS=\{\st,\B\}$, $s=\st$ iff $s'=\st$. Therefore,
$s=s'$.\cqfd
\end{prf}

%% file: sn.tex
%%%%%%%%%%%%%%%%%%%%%%%%%%%%%%%%%%%%%%%%%%%%%%%%%%%%%%%%%%%%%%%%%%%%%%%%%%%%%%
% strong normalization
%%%%%%%%%%%%%%%%%%%%%%%%%%%%%%%%%%%%%%%%%%%%%%%%%%%%%%%%%%%%%%%%%%%%%%%%%%%%%%

\section{Strong normalization}
\label{sec-sn}

Let $\SN$ (resp. $\WN$) be the set of strongly (resp. weakly)
normalizable terms, and $t\nf$ be the normal form of a term $t\in\WN$
($\a$ is assumed confluent).

%%%%%%%%%%%%%%%%%%%%%%%%%%%%%%%%%%%%%%%%%%%%%%%%%%%%%%%%%%%%%%%%%%%%%%%%%%%%%%
% reducibility candidates
%%%%%%%%%%%%%%%%%%%%%%%%%%%%%%%%%%%%%%%%%%%%%%%%%%%%%%%%%%%%%%%%%%%%%%%%%%%%%%

\begin{dfn}[Reducibility candidates]
We assume given a set $\CT$ of {\em constructor terms}.\footnote{$\CT$
is defined in Definition \ref{def-cons}.} A term $t$ is {\em neutral}
if it is not an abstraction, not a constructor term, nor of the form
$f\vt$ with $f\in\DF$ and $|\vt|<|\vl|$ for some rule $f\vl\a
r\in\cR$. We inductively define the set $\cR_t$ of the interpretations
for the terms of type $t$, the ordering $\le_t$ on $\cR_t$, the
element $\top_t\in\cR_t$, and the functions $\biget_t$ and $\bigou_t$
from the powerset of $\cR_t$ to $\cR_t$ as follows. If
$t\notin\cK\cup\{\B\}$ then:

\begin{lst}{--}
\item $\cR_t=\{\vide\}$, $\le_t=\sle$ and $\biget_t(\Re)= \bigou_t(\Re)=
\top_t=\vide$.
\end{lst}

\noindent
Otherwise:

\begin{lst}{--}
\item $\cR_s$ is the set of all the subsets $R$ of $\cT$ such that:
\begin{bfenumii}{R}
\item $R\sle\SN$ (strong normalization).
\item If $t\in R$ then $\a\!\!(t)\sle R$ (stability by reduction).
\item If $t$ is neutral and $\a\!\!(t)\sle R$ then $t\in R$ (neutral terms).
\end{bfenumii}
Furthermore, $\le_s=\sle$, $\top_s=\SN$, $\bigou_s(\Re)=\bigcup\Re$,
$\biget_s(\Re)=\bigcap\Re$ if $\Re\neq\vide$, and $\biget_s(\vide)=
\top_s$.

\item $\cR_{(x:U)K}$ is the set of functions $R$ from $\cT\times\cR_U$
to $\cR_K$ such that $R(u,S)=R(u',S)$ whenever $u\a u'$ or
$\u{u}=\u{u'}$, $\top_{(x:U)K}(u,S)=\top_K$,
$\biget_{(x:U)K}(\Re)(u,S)= \biget_K(\{R(u,S)~|~R\in\Re\})$,
$\bigou_{(x:U)K}(\Re)(u,S)= \bigou_K(\{R(u,S)~|~R\in\Re\})$ and
$R\le_{(x:U)K} R'$ iff $R(u,S)\le_K R'(u,S)$.
\end{lst}

\noindent
Let $(\vt,\vS)\le_i (\vt',\vS')$ iff $\vt=\vt'$, $S_i\le S_i'$ and,
for all $j\neq i$, $S_j=S_j'$. A function $R\in\cR_{(\vx:\vT)\st}$ is
{\em monotone} (resp. {\em anti-monotone}) in its $i$th argument if
$R(\vQ)\le R(\vQ')$ whenever $\vQ\le_i \vQ'$ (resp. $\vQ\ge_i
\vQ'$). Let $\cR_\tf^m$ be the set of functions $R\in\cR_\tf$ such
that $R$ is monotone in all its arguments $i\in\mon^+(f)$, and
anti-monotone in all its arguments $i\in\mon^-(f)$.
\end{dfn}

%%%%%%%%%%%%%%%%%%%%%%%%%%%%%%%%%%%%%%%%%%%%%%%%%%%%%%%%%%%%%%%%%%%%%%%%%%%%%%
% properties
%%%%%%%%%%%%%%%%%%%%%%%%%%%%%%%%%%%%%%%%%%%%%%%%%%%%%%%%%%%%%%%%%%%%%%%%%%%%%%

\begin{lem}
\label{lem-cand}
$(\cR_t,\le_t)$ and $(\cR_t^m,\le_t)$ are complete lattices with
$\top_t$ as their greatest element and $\biget_t(\Re)$ as the greatest
lower bound of $\Re$. Moreover:

\begin{lst}{--}
\item If $\Re$ is totally ordered then $\bigou_t(\Re)$ is the lowest
upper bound of $\Re$.
\item For all $R\in\cR_s$, $\cX\sle R$.
\item If $\G\th t:T$ and $\t:\G\leadsto\D$ then $\cR_{T\t}=\cR_T$.
\item If $\G\th t:T$ then $\cR_{T\vphi}=\cR_T$.
\item The smallest element $\bot_s=\biget_s(\cR_s)$ only contains
neutral terms.
\end{lst}
\end{lem}

\begin{prf}
The proof is similar to the one for CAC \cite{blanqui03mscs}.\cqfd
\end{prf}

%%%%%%%%%%%%%%%%%%%%%%%%%%%%%%%%%%%%%%%%%%%%%%%%%%%%%%%%%%%%%%%%%%%%%%%%%%%%%%
% candidates and subtyping
%%%%%%%%%%%%%%%%%%%%%%%%%%%%%%%%%%%%%%%%%%%%%%%%%%%%%%%%%%%%%%%%%%%%%%%%%%%%%%

\begin{lem}
If $\G\th T\le T':s$ then $\cR_T=\cR_{T'}$.
\end{lem}

\begin{prf}
If $s=\st$ then $\cR_T=\{\vide\}=\cR_{T'}$. Assume now that $s=\B$.
We proceed by induction on $T\le T'$.

\begin{lst}{}
\item [\bf(refl)] Immediate.

\item [\bf(symb)] Not possible.

\item [\bf(prod)] $\cR_{(x:U)V}$ is the set of functions from
$\cT\times\cR_U$ to $\cR_V$ that are invariant by reduction and size
substitution. $\cR_{(x:U')V'}$ is the set of functions from
$\cT\times\cR_{U'}$ to $\cR_{V'}$ that are invariant by reduction and
size substitution. By induction hypothesis, $\cR_U=\cR_{U'}$ and
$\cR_V=\cR_{V'}$. Therefore, $\cR_{(x:U)V}= \cR_{(x:U')V'}$.

\item [\bf(conv)] By induction hypothesis,
$\cR_{T'}=\cR_{U'}$. Since $\cR_T=\cR_{T'}$ and $\cR_U=\cR_{U'}$, we
have $\cR_T=\cR_U$.\cqfd
\end{lst}
\end{prf}

%%%%%%%%%%%%%%%%%%%%%%%%%%%%%%%%%%%%%%%%%%%%%%%%%%%%%%%%%%%%%%%%%%%%%%%%%%%%%%
% interpretation schema
%%%%%%%%%%%%%%%%%%%%%%%%%%%%%%%%%%%%%%%%%%%%%%%%%%%%%%%%%%%%%%%%%%%%%%%%%%%%%%

\begin{dfn}[Interpretation schema]
\label{def-schema-int}
A {\em candidate assignment} is a function $\xi$ from $\cX$ to
$\bigcup \,\{\cR_t ~|~ t\in\cT\}$. A candidate assignment $\xi$ {\em
validates} an environment $\G$ or is a $\G$-assignment,
$\xi\models\G$, if, for all $x\in\dom(\G)$, $x\xi\in \cR_{x\G}$.

An {\em interpretation} for a symbol $C\in\CFB$ is a monotone function
$I$ from $\fA$ to $\cR_\tf^m$. An {\em interpretation} for a symbol
$f\notin\CFB$ is an element of $\cR_\tf^m$. An {\em interpretation}
for a set $\cG$ of predicate symbols is a function which, to every
symbol $g\in\cG$, associates an interpretation for $g$.

The {\em interpretation} of $t$ w.r.t. a candidate assignment $\xi$,
an interpretation $I$ for $\cF$, a substitution $\t$ and a valuation
$\nu$, $\I{t}_{\xi,\t}^{I,\nu}$, is defined by induction on $t$:

\begin{lst}{--}
\item $\I{t}^{I,\nu}_{\xi,\t}= \top_t$ if $t\in\cO\cup\cS$
\item $\I{F}^{I,\nu}_{\xi,\t}= I_F$ if $F\in\DFB$
\item $\I{C^a}^{I,\nu}_{\xi,\t}= I_C^{a\nu}$ if $C\in\CFB$
\item $\I{x}^{I,\nu}_{\xi,\t}= x\xi$
\item $\I{(x:U)V}^{I,\nu}_{\xi,\t}= \{t\in\cT~|~ \all
u\in\I{U}^{I,\nu}_{\xi,\t}, \all S\in\cR_U,
tu\in\I{V}^{I,\nu}_{\xi_x^S,\t_x^u}\}$
\item $\I{[x:U]v}^{I,\nu}_{\xi,\t}(u,S)=
\I{v}^{I,\nu}_{\xi_x^S,\t_x^u}$
\item $\I{tu}^{I,\nu}_{\xi,\t}=
\I{t}^{I,\nu}_{\xi,\t}(u\t,\I{u}^{I,\nu}_{\xi,\t})$
\end{lst}

\noindent
where $\t_x^u=\t\cup\xu$ and $\xi_x^S=\xi\cup\xS$.

Let $I$ be an interpretation for $\cF$. A symbol $f$ is {\em
computable} if, for all $\nu$, $f\in\I\tf^{I,\nu}$. A substitution
$\t$ is {\em adapted} to a $\G$-assignment $\xi$ and a valuation
$\nu$, $\xi,\t\models_\nu\G$, if $\dom(\t)\sle\dom(\G)$ and, for all
$x\in\dom(\t)$, $x\t\in \I{x\G}^{I,\nu}_{\xi,\t}$. The interpretation
is {\em invariant by reduction} if, for all $\nu,\xi,\t$ and
$t,t'\in\WN$, $\I{t}^{I,\nu}_{\xi,\t}= \I{t'}^{I,\nu}_{\xi,\t}$
whenever $t\a t'$.
\end{dfn}

%%%%%%%%%%%%%%%%%%%%%%%%%%%%%%%%%%%%%%%%%%%%%%%%%%%%%%%%%%%%%%%%%%%%%%%%%%%%%%
% interpretations are candidates
%%%%%%%%%%%%%%%%%%%%%%%%%%%%%%%%%%%%%%%%%%%%%%%%%%%%%%%%%%%%%%%%%%%%%%%%%%%%%%

\begin{lem}
\begin{lst}{--}
\item If $\G\th t:T$ and $\xi\models\G$ then $\I{t}_{\xi,\t}^{I,\nu}\in\cR_T$.
\item If $\t\a\t'$ or $\u\t=\u\t'$ then $\I{t}_{\xi,\t}^{I,\nu}=
\I{t}_{\xi,\t'}^{I,\nu}$.
\end{lst}
\end{lem}

\begin{prf}
The proof is similar to the one for CAC \cite{blanqui03mscs}.\cqfd
\end{prf}

%%%%%%%%%%%%%%%%%%%%%%%%%%%%%%%%%%%%%%%%%%%%%%%%%%%%%%%%%%%%%%%%%%%%%%%%%%%%%%
% substitution lemma for candidates
%%%%%%%%%%%%%%%%%%%%%%%%%%%%%%%%%%%%%%%%%%%%%%%%%%%%%%%%%%%%%%%%%%%%%%%%%%%%%%

\begin{lem}[Candidate substitution]
\label{lem-cand-subs}
If $\G\th t:T$, $\g:\G\leadsto\D$ and $\xi\models\D$ then
$\I{t\g}^{I,\nu}_{\xi,\s}= \I{t}^{I,\nu}_{\eta,\g\s}$ with $x\eta=
\I{x\g}^{I,\nu}_{\xi,\s}$ and $\eta\models\G$.
\end{lem}

\begin{prf}
The proof is similar to the one for CAC \cite{blanqui03mscs}.\cqfd
\end{prf}

%%%%%%%%%%%%%%%%%%%%%%%%%%%%%%%%%%%%%%%%%%%%%%%%%%%%%%%%%%%%%%%%%%%%%%%%%%%%%%
% substitution lemma for size expressions
%%%%%%%%%%%%%%%%%%%%%%%%%%%%%%%%%%%%%%%%%%%%%%%%%%%%%%%%%%%%%%%%%%%%%%%%%%%%%%

\begin{lem}[Size substitution]
\label{lem-size-subs}
If $\G\th t:T$ then $\I{t\vphi}^{I,\nu}_{\xi,\t}=
\I{t}^{I,\vphi\nu}_{\xi,\t}$ where $\alpha(\vphi\nu)=
(\alpha\vphi)\nu$.
\end{lem}

\begin{prf}
By induction on $t$.
\begin{lst}{--}
\item If $t$ is an object, a sort or a symbol $f\in\Fs$ then $t\vphi$
is of the same kind and $\I{t\vphi}^{I,\nu}_{\xi,\t}=
\I{t\vphi}^{I,\nu}_{\xi,\t}= \top_t$.

\item $\I{C^a\vphi}^{I,\nu}_{\xi,\t}= I_C^{a\vphi\nu}=
\I{C^a}^{I,\vphi\nu}_{\xi,\t}$.

\item $\I{x\vphi}^{I,\nu}_{\xi,\t}= \I{x}^{I,\nu}_{\xi,\t}= x\xi$.

\item $\I{(x:U\vphi)V\vphi}^{I,\nu}_{\xi,\t}= \{t\in\cT~|~ \all
u\in\I{U\vphi}^{I,\nu}_{\xi,\t}, \all S\in\cR_{U\vphi},
tu\in\I{V\vphi}^{I,\nu}_{\xi_x^S,\t_x^u}\}$. By induction hypothesis,
$\I{U\vphi}^{I,\nu}_{\xi,\t}= \I{U}^{I,\vphi\nu}_{\xi,\t}$ and
$\I{V\vphi}^{I,\nu}_{\xi_x^S,\t_x^u}=
\I{V}^{I,\vphi\nu}_{\xi_x^S,\t_x^u}$. And since $\cR_{U\vphi}= \cR_U$,
$\I{(x:U\vphi)V\vphi}^{I,\nu}_{\xi,\t}= \I{(x:U)V}^{I,\nu}_{\xi,\t}$.

\item If $\G\th [x:U]v:T$ then, by inversion, $\G\th [x:U]v:(x:U)V$
for some $V$, and $\G\vphi\th
[x:U\vphi]v\vphi:(x:U\vphi)V\vphi$. Since $\cR_{U\vphi}=\cR_U$ and
$\cR_{V\vphi}=\cR_V$, $\I{[x:U\vphi]v\vphi}^{I,\nu}_{\xi,\t}$ has the
same domain and codomain as
$\I{[x:U]v}^{I,\nu}_{\xi,\t}$. Furthermore,
$\I{[x:U\vphi]v\vphi}^{I,\nu}_{\xi,\t}(u,S)=
\I{v\vphi}^{I,\nu}_{\xi_x^S,\t_x^u}= \I{v}^{I,\nu}_{\xi_x^S,\t_x^u}$
by induction hypothesis.

\item $\I{t\vphi u\vphi}^{I,\nu}_{\xi,\t}=
\I{t\vphi}^{I,\nu}_{\xi,\t}(u\vphi\t,\I{u\vphi}^{I,\nu}_{\xi,\t})=
\I{t}^{I,\vphi\nu}_{\xi,\t}(u\t,\I{u}^{I,\vphi\nu}_{\xi,\t})$
by induction hypothesis and invariance by size change.\cqfd\\
\end{lst}
\end{prf}

%%%%%%%%%%%%%%%%%%%%%%%%%%%%%%%%%%%%%%%%%%%%%%%%%%%%%%%%%%%%%%%%%%%%%%%%%%%%%%
% positive and negative positions
%%%%%%%%%%%%%%%%%%%%%%%%%%%%%%%%%%%%%%%%%%%%%%%%%%%%%%%%%%%%%%%%%%%%%%%%%%%%%%

We now define the sets of positive and negative positions in a term,
which will enforce monotony and anti-monotony properties
respectively.

\begin{dfn}[Positive and negative positions]
The set of positions in a term $t$ is inductively defined as
follows:\footnote{It is defined so that $\pos(\u{t})\sle\pos(t)$.}

\begin{lst}{--}
\item $\pos(s)= \pos(x)= \pos(f)= \{\vep\}$
\item $\pos((x:u)v)= \pos([x:u]v)= \pos(uv)= 1.\pos(u)\cup 2.\pos(v)$
\item $\pos(C^a)= \{\vep\}\cup 0.\pos(a)$
\end{lst}

Let $\pos(x,t)$ be the set of positions of the free occurrences of $x$
in $t$, and $\pos(f,t)$ be the set of positions of the occurrences of
$f$ in $t$. The set of {\em positive positions} in $t$, $\pp(t)$, and
the set of {\em negative positions} in $t$, $\pm(t)$, are
simultaneously defined by induction on $t$:

\begin{lst}{--}
\item $\pd(s)= \pd(x)= \{\vep~|~\d=+\}$
\item $\pd((x:U)V)= 1.\pmd(U)\cup 2.\pd(V)$
\item $\pd([x:U]v)= 2.\pd(v)$
\item $\pd(tu)= 1.\pd(t)$ if $t\neq f\vt$
\item $\pd(f\vt)= \{1^{|\vt|}~|~\d=+\}\cup
  \,\bigcup\{1^{|\vt|-i}2.\pos^{\vep\d}(t_i)~|~ \vep\in\{-,+\},
  i\in\mon^\vep(f)\}$
\item $\pd(C^a\vt)= \pd(C\vt)\cup  \{1^{|\vt|}0~|~\d=+\}.\pd(a)$.
\end{lst}

\noindent
where $\d\in\{-,+\}$, $-+=-$ and $--=+$ (usual rule of signs).
\end{dfn}

%%%%%%%%%%%%%%%%%%%%%%%%%%%%%%%%%%%%%%%%%%%%%%%%%%%%%%%%%%%%%%%%%%%%%%%%%%%%%%
% monotony
%%%%%%%%%%%%%%%%%%%%%%%%%%%%%%%%%%%%%%%%%%%%%%%%%%%%%%%%%%%%%%%%%%%%%%%%%%%%%%

\begin{lem}[Monotony]
\label{lem-mon}
Let $\le^+=\le$; $\le^-=\ge$; $\xi\le_x\xi'$ iff $x\xi\le x\xi'$ and,
for all $y\neq x$, $y\xi=y\xi'$; $I\le_f I'$ iff $I_f\le I_f'$ and,
for all $g\neq f$, $I_g=I_g'$; $\nu\le_\alpha\nu'$ iff
$\alpha\nu\le_\fA\alpha\nu'$ and, for all $\b\neq\alpha$,
$\b\nu=\b\nu'$. Assume that $\G\th t:T$ and $\xi,\xi'\models\G$.

\begin{lst}{--}
\item If $\xi\le_x\xi'$ and $\pos(x,t)\sle\pd(t)$ then
$\I{t}_{\xi,\t}^{I,\nu}\le^\d \I{t}_{\xi',\t}^{I,\nu}$.

\item If $I\le_f I'$ and $\pos(f,t)\sle\pd(t)$ then
$\I{t}_{\xi,\t}^{I,\nu}\le^\d \I{t}_{\xi,\t}^{I',\nu}$.

\item If $\nu\le_\alpha\nu'$ and $\pos(\alpha,t)\sle\pd(t)$
then $\I{t}_{\xi,\t}^{I,\nu}\le^\d \I{t}_{\xi,\t}^{I,\nu'}$.

\item If $\G\th T\le T':s$, $T,T'\in\WN$ and the interpretation
is invariant by reduction then $\I{T}_{\xi,\t}^{I,\nu}\le
\I{T'}_{\xi,\t}^{I,\nu}$.
\end{lst}
\end{lem}

\begin{prf}
\begin{lst}{--}
\item The first two properties are proved for CAC in
\cite{blanqui03mscs} and their proofs are still valid.

\item We now prove the third property. It uses the same techniques.
So, we only detail the case $t=C^a\vt$. Let $R=
\I{t}_{\xi,\t}^{I,\nu}$ and $R'= \I{t}_{\xi,\t}^{I,\nu'}$. $R=
I^{a\nu}_C(\vt\t,\vS)$ with $\vS= \I{\vt}^{I,\nu}_{\xi,\t}$, and $R'=
I^{a\nu'}_C(\vt\t,\vS')$ with $\vS= \I{\vt}^{I,\nu'}_{\xi,\t}$. Let
$n=|\vt|$ and $i\in\{1,\ldots,n\}$. If $\pos(\alpha,t_i)=\vide$ then
$S_i=S_i'$. Otherwise, since $\pos(\alpha,t)\sle \pd(t)$, there is
$\vep_i$ such that $i\in\mon^{\vep_i}(f)$ and $\pos(\alpha,t_i)\sle
\pos^{\vep_i\d}(t_i)$. Thus, by induction hypothesis, $S_i\le^{\vep_i\d}
S_i'$. Let $Q^k_j= (\vt\t,S'_j)$ if $j\le k$, and $Q^k_j= (\vt\t,S_j)$
if $j>k$. We have $\vQ^0=(\vt\t,\vS)$, $\vQ^n=(\vt\t,\vS')$ and, for
all $k\in \{1,\ldots,n\}$, $\vQ^{k-1} \le_k^{\vep_k\d} \vQ^k$. Thus,
$I^{a\nu}_C(\vQ^{k-1}) \le^{\vep_k^2\d} I^{a\nu}_C(\vQ^k)$, that is,
$I^{a\nu}_C(\vQ^{k-1}) \le^\d I^{a\nu}_C(\vQ^k)$ since $\vep_k^2=+$
and symbol interpretations are monotone in their monotone arguments
and anti-monotone in their anti-monotone arguments. So, $R=
I^{a\nu}_C(\vQ^0)\le^\d I^{a\nu}_C(\vQ^n)$. Now, if
$\pos(\alpha,C^a)=\vide$ then $a\nu=a\nu'$ and $R\le^\d R'=
I^{a\nu}_C(\vQ^n)$. Otherwise, $\d=+$ and $a\nu\le_\fA a\nu'$ since
$\pos(\alpha,a)\sle\pp(a)$. Thus, $R\le R'$ since symbol
interpretations are monotone functions on $\fA$.

\item We now prove the last property by induction on $T\le T'$. Let $R=
\I{T}_{\xi,\t}^{I,\nu}$ and $R'= \I{T'}_{\xi,\t}^{I,\nu}$,

\begin{lst}{}
\item [\bf(refl)] Immediate.
  
\item [\bf(symb)] Let $\vQ= (\vt\t,\I{\vt}^{I,\nu}_{\xi,\t})$. We
have $R= I^{a\nu}_C(\vQ)\le R'= I^{b\nu}_C(\vQ)$ since $a\nu\le_\fA
b\nu$ and symbol interpretations are monotone on $\fA$.

\item [\bf(prod)] Let $t\in R$, $u\in
\I{U'}^{I,\nu}_{\xi,\t}$ and $S\in\cR_{U'}$. We must prove that $tu\in
\I{V'}^{I,\nu}_{\xi_x^S,\t_x^u}$. By induction hypothesis,
$\I{U'}_{\xi,\t}^{I,\nu}\le \I{U}_{\xi,\t}^{I,\nu}$. So, $u\in
\I{U}_{\xi,\t}^{I,\nu}$. Since $\cR_{U'}=\cR_U$ and $t\in R$, $tu\in
\I{V}_{\xi_x^S,\t_x^u}^{I,\nu}$. Now, by induction hypothesis,
$\I{V}_{\xi_x^S,\t_x^u}^{I,\nu}\le
\I{V'}_{\xi_x^S,\t_x^u}^{I,\nu}$. Therefore, $tu\in
\I{V'}_{\xi_x^S,\t_x^u}^{I,\nu}$.

\item [\bf(conv)] By induction hypothesis,
$\I{T'}_{\xi,\t}^{I,\nu}\le \I{U'}_{\xi,\t}^{I,\nu}$. Since
$T,U\in\WN$ and the interpretation is invariant by reduction,
$\I{T'}_{\xi,\t}^{I,\nu}=R$ and $\I{U'}_{\xi,\t}^{I,\nu}=
R'$. Therefore, $R\le R'$.\cqfd
\end{lst}
\end{lst}
\end{prf}

%%%%%%%%%%%%%%%%%%%%%%%%%%%%%%%%%%%%%%%%%%%%%%%%%%%%%%%%%%%%%%%%%%%%%%%%%%%%%%
% strong normalization
%%%%%%%%%%%%%%%%%%%%%%%%%%%%%%%%%%%%%%%%%%%%%%%%%%%%%%%%%%%%%%%%%%%%%%%%%%%%%%

\begin{thm}[Strong normalization]
\label{thm-cond-sn}
If there is an interpretation $I$ invariant by reduction and such that
every symbol is computable then every well-typed term is strongly
normalizable.
\end{thm}

\begin{prf}
One first prove by induction that, if $\G\th t:T$ then, for all $\xi$,
$\nu$ and $\t$ such that $\xi\models\G$ and $\xi,\t\models_\nu\G$,
then $t\t\in \I{T}_{\xi,\t}^\nu$. Then, one prove that, if $x\t=x$ and
$x\xi=\top_{x\G}$, then $\xi\models\G$ and $\xi,\t\models_\nu\G$. See
\cite{blanqui03mscs} for details.\cqfd
\end{prf}

%% file: cons.tex
%%%%%%%%%%%%%%%%%%%%%%%%%%%%%%%%%%%%%%%%%%%%%%%%%%%%%%%%%%%%%%%%%%%%%%%%%%%%%%
% constructor-based systems
%%%%%%%%%%%%%%%%%%%%%%%%%%%%%%%%%%%%%%%%%%%%%%%%%%%%%%%%%%%%%%%%%%%%%%%%%%%%%%

\section{Constructor-based systems}
\label{sec-cons}

We now study the case of CACSA's whose size algebra contains the
following expressions (at least):

\begin{center}
$a ::= \alpha ~|~ sa ~|~ \infty ~|~ \ldots$
\end{center}

In case that there is no other symbol, the ordering $\le_\cA$ on size
expressions is defined as the smallest quasi-ordering $\le$ such that,
for all $a$, $a<sa$ and $a\le\infty$. We interpret size expressions in
the set $\fA=\W+1$, where $\W$ is the first uncountable ordinal, by
taking:

\begin{lst}{--}
\item $s_\fA(\fa)= \fa+1$ if $\fa<\W$, and $\W$ otherwise.
\item $\infty_\fA= \W$.
\end{lst}

One can easily imagine other size expressions like $a+b$, $max(a,b)$,
\ldots

%%%%%%%%%%%%%%%%%%%%%%%%%%%%%%%%%%%%%%%%%%%%%%%%%%%%%%%%%%%%%%%%%%%%%%%%%%%%%%
% monotony conditions
%%%%%%%%%%%%%%%%%%%%%%%%%%%%%%%%%%%%%%%%%%%%%%%%%%%%%%%%%%%%%%%%%%%%%%%%%%%%%%

\begin{dfn}[Constructor-based system]
\label{def-cons}
We assume given a {\em precedence} $\le_\cF$ on $\cF$, that is, a
quasi-ordering whose strict part $>_\cF$ is well-founded, and that
every $C\in\CFB$ with $C:(\vz:\vV)\st$ is equipped with a set $\co(C)$
of {\em constructors}, that is, a set of constant symbols
$f:(\vy:\vU)C^a\vv$ equipped with a set $\acc(f)\sle
\{1,\ldots,|\vy|\}$ of {\em accessible} arguments such that:

\begin{lst}{\bu}
\item If there are $D=_\cF C$ and $j\in\acc(c)$ such that
$\pos(D,U_j)\neq\vide$ then $\cV(\tf)=\{\alpha\}$ and $a=s\alpha$.

\item For all $j\in\acc(c)$:
\begin{lst}{--}
\item For all $D>_\cF C$, $\pos(D,U_j)=\vide$.
\item For all $D\simeq_\cF C$ and $p\in\pos(D,U_j)$, $p\in\pp(U_j)$
and $U_j|_p=D^\alpha$.
\item For all $p\in\pos(\alpha,U_j)$, $p=q0$, $U_j|_q=D^\alpha$
and $D\simeq_\cF C$.
\item For all $x\in\FVB(U_j)$, there is $\io_x$ with
$v_{\io_x}=x$ and $\pos(x,U_j)\sle\pp(U_j)$.
\end{lst}

\item For all $F\in\DFB$ and $F\vl\a r\in\cR$:
\begin{lst}{--}
\item For all $G>_\cF F$, $\pos(G,r)=\vide$.
\item For all $i\in\mon^\d(F)$, $l_i\in\XB$ and $\pos(l_i,r)\sle\pd(r)$.
\item For all $x\in\FVB(r)$, there is $\ka_x$ with $l_{k_x}=x$.
\end{lst}
\end{lst}

\noindent
A {\em $C$-constructor term} is a term of the form $f\vu$ with
$f\in\co$, $f:(\vy:\vU)C^a\vv$, $|\vu|=|\vy|$ and
$\acc(f)\neq\vide$. Let $\CT(C)$ be the set of $C$-constructor terms.
\end{dfn}

The conditions involving $\io_x$ and $\ka_x$ means that we restrict
our attention to {\em small} inductive types. Strong elimination, that
is, predicate-level recursion on big inductive types may lead to
non-termination \cite{coquand86lics}. Yet, weak elimination, that is,
object-level recursion on big inductive types is admissible. As shown
in \cite{blanqui03tlca}, it is possible to raise this restriction at
the price of not being allowed to match defined symbols.

%%%%%%%%%%%%%%%%%%%%%%%%%%%%%%%%%%%%%%%%%%%%%%%%%%%%%%%%%%%%%%%%%%%%%%%%%%%%%%
% primitive types
%%%%%%%%%%%%%%%%%%%%%%%%%%%%%%%%%%%%%%%%%%%%%%%%%%%%%%%%%%%%%%%%%%%%%%%%%%%%%%

Among constant predicate symbols, we distinguish the class of
primitive types that includes all first-order data type like natural
numbers, lists of natural numbers, \ldots Primitive types are not
polymophic but they can have primitive dependancies like the type of
arrays of natural numbers.

\begin{dfn}[Primitive types]
\label{def-prim}
A symbol $C\in\CFB$ is {\em primitive} if $\tC=(\vz:\vV)\st$,
$\{\vz\}\sle\Xs$ and, for all $D\simeq_\cF C$, for all constructor
$f:(\vy:\vU)D^{s\alpha}\vv$ and for all $j\in\acc(f)$, either
$U_j=E^\infty\vt$ with $E<_\cF C$ and $E$ primitive, or
$U_j=E^\alpha\vt$ with $E\simeq_\cF C$. The {\em size} of a term $t$
in a primitive type $C$ is defined as follows. If $t$ is a constructor
term $f\vu$ with $f:(\vy:\vU)C^{s\alpha}\vv$ and, for all
$j\in\acc(f)$ such that $\pos(\alpha,U_j)\neq\vide$,
$U_j=C_j^\alpha\vv^j$, then $|t|_C=1+max\{|u_j|_{C_j}~|~j\in\acc(f),
\pos(\alpha,U_j)\neq\vide\}$. Otherwise, $|t|_C=0$.
\end{dfn}

%%%%%%%%%%%%%%%%%%%%%%%%%%%%%%%%%%%%%%%%%%%%%%%%%%%%%%%%%%%%%%%%%%%%%%%%%%%%%%
% interpretation of defined predicate symbols
%%%%%%%%%%%%%%%%%%%%%%%%%%%%%%%%%%%%%%%%%%%%%%%%%%%%%%%%%%%%%%%%%%%%%%%%%%%%%%

We define the interpretation of predicate symbols by induction on
$>_\cF$.

\begin{dfn}[Interpretation of defined predicate symbols]
\hfill Assume\\ that $F:{(\vx:\vT)}U$. We take $I_F(\vt,\vS)=
\I{r}_{\xi,\s}^{I}$ if $\vt\in\WN$, $\vt\nf=\vl\s$, $F\vl\a r\in\cR$
and $x\xi=S_{\ka_x}$. Otherwise, we take $I_F(\vt,\vS)= \top_U$.
\end{dfn}

Thanks to Lemma \ref{lem-mon}, one can easily check that $I$ is
monotone in its monotone arguments. The well-foundedness of the
definition is a consequence of the correctness of the termination
criterion.

%%%%%%%%%%%%%%%%%%%%%%%%%%%%%%%%%%%%%%%%%%%%%%%%%%%%%%%%%%%%%%%%%%%%%%%%%%%%%%
% interpretation of predicate symbols
%%%%%%%%%%%%%%%%%%%%%%%%%%%%%%%%%%%%%%%%%%%%%%%%%%%%%%%%%%%%%%%%%%%%%%%%%%%%%%

We now define the interpretation of a constant predicate symbols
by transfinite induction on $\fa\in\fA$.

\begin{dfn}[Interpretation of constant predicate symbols]
\label{def-int-const}
\hfill
\begin{lst}{--}
\item $I_C^0(\vS)$\footnote{We do not write $\vt$ since the
interpretation does not depend on it.} is the set of $u\in\SN$ such
that $u$ never reduces to a $C$-constructor term.

\item $I_C^{\fa+1}(\vS)$ is the set of terms $u\in\SN$
such that, if $u$ reduces to a constructor term $f\vu$ with
$f:(\vy:\vU)C^{s\alpha}\vv$ then, for all $j\in\acc(f)$, $u_j\in
\I{U_j}^{I,\nu}_{\xi,\t}$ with $y\xi= S_{\io_y}$, $\vy\t=\vu$ and
$\alpha\nu=\fa$.

\item $I_C^\fb= \biget_\tC(\{I_C^\fa~|~\fa<\fb\})$
if $\fb$ is a limit ordinal.
\end{lst}

\noindent
Let $K_C^\fa(\vS)= I_C^\fa(\vS)\cap \CT(C)$ and, for $t\in
I^\W_C(\vS)$, let $o_{C(\vS)}(t)$ be the smallest ordinal $\fa$ such
that $t\in I_C^\fa(\vS)$.
\end{dfn}

The interpretation is well defined thanks to the assumptions made on
$U_j$ when $j$ is accessible.

%%%%%%%%%%%%%%%%%%%%%%%%%%%%%%%%%%%%%%%%%%%%%%%%%%%%%%%%%%%%%%%%%%%%%%%%%%%%%%
% monotony
%%%%%%%%%%%%%%%%%%%%%%%%%%%%%%%%%%%%%%%%%%%%%%%%%%%%%%%%%%%%%%%%%%%%%%%%%%%%%%

\begin{lem}
\label{lem-suc}
If $f\vu\in K^\W_C(\vS)$ then $o_{C(\vS)}(f\vu)$ is a
successor ordinal.
\end{lem}

\begin{prf}
Assume that $\fa=o_{C(\vS)}(f\vu)$ is a limit ordinal. Then,
$I^\fa_C(\vS)= \bigcup \{I^\fb_C(\vS)~|$ $\fb<\fa\}$ and $t\s\in
I^\fb_C(\vS)$ for some $\fb<\fa$, which is not possible. Now, $\fa\neq
0$ since $K^0_C(\vS)= \vide$. Therefore, $\fa$ is a successor
ordinal.\cqfd
\end{prf}

%%%%%%%%%%%%%%%%%%%%%%%%%%%%%%%%%%%%%%%%%%%%%%%%%%%%%%%%%%%%%%%%%%%%%%%%%%%%%%

\begin{lem}
$I$ is monotone.
\end{lem}

\begin{prf}
We prove that $\fa\le\fb\A I^\fa\le I^\fb$ by induction on $\fa$.

\begin{lst}{\bu}
\item $\fa=0$. 
\begin{lst}{--}
\item $\fb=0$. Immediate.
\item $\fb=\fb'+1$. By induction hypothesis, $I^0\le I^{\fb'}$.
We now prove that $I^{\fb'}\le I^{\fb'+1}$. Let $t\in
I^{\fb'}_C(\vS)$. Then, $t\in\SN$. Assume now that $t$ reduces to a
constructor term $f\vu$ with $f:(\vy:\vU)C^{s\alpha}\vv$. By Lemma
\ref{lem-suc}, $t\in I^{\fc+1}_C(\vS)$ for some $\fc<\fb'$. Let
$j\in\acc(f)$. Then, $u_j\in \I{U_j}^\nu_{\xi,\t}$ with $y\xi=
S_{\io_y}$, $\vy\t=\vu$ and $\alpha\nu=\fc$. After the conditions on
$U_j$, by Lemma \ref{lem-mon}, $\I{U_j}^\nu_{\xi,\t}\sle
\I{U_j}^\mu_{\xi,\t}$ where $\alpha\mu=\fb'$. Thus, $t\in
I^{\fb'+1}_C(\vS)$.
\item $\fb$ is a limit ordinal. By induction hypothesis,
$I^0\le I^{\fb'}$ for all $\fb'<\fb$. Thus, $I^0\le I^\fb$.
\end{lst}

\item $\fa=\fa'+1$.
\begin{lst}{--}
\item $\fb=0$. Not possible.
\item $\fb=\fb'+1$. Then, $\fa'\le\fb'$. Let $t\in I^\fa_C(\vS)$. Then,
$t\in\SN$. Assume now that $t$ reduces to a constructor term $f\vu$
with $f:(\vy:\vU)C^{s\alpha}\vv$ and let $j\in\acc(f)$. Then, $u_j\in
\I{U_j}^\nu_{\xi,\t}$ with $y\xi= S_{\io_y}$, $\vy\t=\vu$ and
$\alpha\nu=\fa'$. After the conditions on $U_j$, by Lemma
\ref{lem-mon}, $\I{U_j}^\nu_{\xi,\t}\sle \I{U_j}^\mu_{\xi,\t}$ where
$\alpha\mu=\fb'$. Thus, $t\in I^\fb_C(\vS)$.
\item $\fb$ is a limit ordinal. Then, $\fa'<\fb'$ for some $\fb'<\fb$
and we can conclude by induction hypothesis.
\end{lst}

\item $\fa$ is a limit ordinal.
\begin{lst}{--}
\item $\fb=0$. Not possible.
\item $\fb=\fb'+1$. Then, $\fa\le\fb'$ and we can conclude by induction
hypothesis.
\item $\fb$ is a limit ordinal. Then, for all $\fa'<\fa$, $\fa'<\fb$, and
we can conclude by induction hypothesis.\cqfd
\end{lst}
\end{lst}
\end{prf}

%%%%%%%%%%%%%%%%%%%%%%%%%%%%%%%%%%%%%%%%%%%%%%%%%%%%%%%%%%%%%%%%%%%%%%%%%%%%%%
% first-order types
%%%%%%%%%%%%%%%%%%%%%%%%%%%%%%%%%%%%%%%%%%%%%%%%%%%%%%%%%%%%%%%%%%%%%%%%%%%%%%

\begin{lem}[Primitive types]
\label{lem-prim}
Let $C$ be primitive type. If $\fa\ge\w$ then $I_C^\fa=
\top_\tC$. Otherwise, $I_C^\fa(\vS)= \{t\in\SN~|~ |t\nf\!|_C\le\fa\}$,
that is, $o_{C(\vs)}(t)=|t\nf\!|_C$.
\end{lem}

\begin{prf}
We proceed by induction on $C$ with $>_\cF$ as well-founded ordering.

Let $J_C^\fa= \{t\in\SN~|~ |t\nf\!|_C\le\fa\}$. Since primitive types
are not polymorphic, every $S_i=\vide$. So, we can drop the arguments
$\vS$. Note also that $|t|_C\le|t'|_C$ whenever $t\a t'$ (since
$\co\sle\CF$).

We first prove that, for all $\fa<\w$, if $o_C(t)=\fa$ then
$|t\nf\!|_C=\fa$.

\begin{lst}{--}
\item $\fa=0$. If $o_C(t)=0$ then $t\in I_C^0\sle J_C^0$. Thus,
$|t\nf\!|_C=0$.

\item $\fa=\fa'+1$. If $o_C(t)=\fa'+1$ then $t\in I_C^{\fa'+1}\moins
I_C^{\fa'}$. Since $t\notin I_C^0$, $t$ reduces to a constructor term
$f\vu$ with $f:(\vy:\vU)C^{s\alpha}\vv$. Let $j\in\acc(f)$. Then,
$u_j\in \I{U_j}^\nu_{\xi,\t}$ with $y\xi= S_{\io_y}$, $\vy\t=\vu$ and
$\alpha\nu=\fa'$. Moreover, either $U_j=C_j^\alpha\vv^j$ with
$C_j\simeq_\cF C$, or $U_j=C_j^\infty\vv^j$ with $C_j<_\cF C$. In the
former case, $u_j\in I_{C_j}^{\fa'}$. Thus, $o_{C_j}(u_j)\le\fa'$ and,
by induction hypothesis, $o_{C_j}(u_j)= |u_j\nf\!|_{C_j}$. Therefore,
$o_C(t)=|t\nf\!|$.
\end{lst}

Thus $o_C(t)=|t\nf\!|_C$ and, for all $\fa<\w$, $I_C^\fa=J_C^\fa$. We
now prove that $I^{\w+1}_C=I^\w_C=\SN$. Let $t\in I^{\w+1}_C\moins
I^\w_C$. Since $t\notin I^0_C$, $t$ reduces to a constructor term
$f\vu$ with $f:(\vy:\vU)C^{s\alpha}\vv$ and, for all $j\in\acc(f)$,
$u_j\in \I{U_j}^\nu_{\xi,\t}$ with $y\xi= S_{\io_y}$, $\vy\t=\vu$ and
$\alpha\nu=\w$. Thus, for all $j\in\acc(f)$, there is $\fa_j<\w$ such
that $u_j\in \I{U_j}^{\nu_j}_{\xi,\t}$ with
$\alpha\nu_j=\fa_j$. $\fa=max\{\fa_j~|~j\in\acc(f)$ is well defined
since $\acc(f)\neq\vide$ and $\fa<\w$ since $\acc(f)$ is finite. Thus,
$t\in I^{\fa+1}_C\sle I^\w_C$.\cqfd\\
\end{prf}

%%%%%%%%%%%%%%%%%%%%%%%%%%%%%%%%%%%%%%%%%%%%%%%%%%%%%%%%%%%%%%%%%%%%%%%%%%%%%%
% computability closure
%%%%%%%%%%%%%%%%%%%%%%%%%%%%%%%%%%%%%%%%%%%%%%%%%%%%%%%%%%%%%%%%%%%%%%%%%%%%%%

We now give general conditions for every symbol to be computable,
based on the fundamental notion of {\em computability closure}. The
computability closure of a term $t$ is a set of terms that can be
proved computable whenever $t$ is computable. If, for every rule
$f\vl\a r$, $r$ belongs to the computability closure of $\vl$, then
rules preserve computability, hence strong normalization.

In \cite{blanqui03mscs}, the computability closure is inductively
defined as a typing relation $\thc$ similar to $\th$ except for the
(symb) case which is replaced by two new cases: (symb$^<$) for symbols
strictly smaller than $f$, and (symb$^=$) for symbols equivalent to
$f$ whose arguments are structurally smaller than $\vl$.

Here, we propose to add a new case for symbols equivalent to $f$ whose
arguments have sizes strictly smaller than those of $\vl$. For
comparing the sizes, one can use metrics like in \cite{xi02hosc}.

%%%%%%%%%%%%%%%%%%%%%%%%%%%%%%%%%%%%%%%%%%%%%%%%%%%%%%%%%%%%%%%%%%%%%%%%%%%%%%
% ordering on symbol arguments
%%%%%%%%%%%%%%%%%%%%%%%%%%%%%%%%%%%%%%%%%%%%%%%%%%%%%%%%%%%%%%%%%%%%%%%%%%%%%%

\begin{dfn}[Ordering on symbol arguments]
\label{def-arg-ord}
For every symbol $f:(\vx:\vT)U$, we assume given two well-founded
domains, $(D_f^\cA,>_f^\cA)$ and $(D_f^\fA,>_f^\fA)$, and two
measure/metric functions $\z_f^\cA:\cA^n\a D_f^\cA$ and
$\z_f^\fA:\fA^n\a D_f^\fA$ ($n=|\vx|$) such that
$(D_f^X,>_f^X)=(D_g^X,>_f^X)$ ($X\in\{\cA,\fA\}$) whenever
$f\simeq_\cF g$, and we define:

\begin{lst}{--}
\item $a_f^i=a$ if $T_i=C^a\vv$, and $a_f^i=\infty$ otherwise.
\item $(f,\vphi)>^\cA(g,\psi)$ iff $f>_\cF g$ or $f\simeq_\cF g$ and
$\z_f^\cA(\va_f\vphi) >_f^\cA \z_g^\cA(\va_g\psi)$.
\item $(f,\nu)>^\fA(g,\mu)$ iff $f>_\cF g$ or $f\simeq_\cF g$ and
$\z_f^\fA(\va_f\nu) >_f^\fA \z_g^\fA(\va_g\mu)$.
\end{lst}

\noindent
Then, we assume that $>^\cA$ is decidable and that (for all $\nu$)
$(f,\vphi\nu)>^\fA (g,\psi\nu)$ whenever $(f,\vphi)>^\cA (g,\psi)$.
\end{dfn}

%%%%%%%%%%%%%%%%%%%%%%%%%%%%%%%%%%%%%%%%%%%%%%%%%%%%%%%%%%%%%%%%%%%%%%%%%%%%%%
% status
%%%%%%%%%%%%%%%%%%%%%%%%%%%%%%%%%%%%%%%%%%%%%%%%%%%%%%%%%%%%%%%%%%%%%%%%%%%%%%

\begin{expl}[Lexicographic and multiset status]
A simple metric is given by assigning a {\em status} to every symbol,
that is, a non-empty sequence of finite multisets of strictly positive
integers, describing a simple combination of lexicographic and
multiset comparisons. Given a set $D$ and a status $\z$ of arity $n$
(biggest integer occurring in it), we define $\I{\z}_D$ on $D^n$ as
follows:

\begin{lst}{--}
\item $\I{M_1\ldots M_k}_D(\vx)= (\I{M_1}_D^m(\vx), \ldots, \I{M_k}_D^m(\vx))$
\item $\I{\{i_1,\ldots,i_p\}}_D^m(\vx)= \{x_{i_1},\ldots,x_{i_p}\}$ (multiset)
\end{lst}

\noindent Now, take $\z_f^X=\I{\z_f}_X$, $D_f^X=\z_f^X(X^n)$ and
$>_f^X=((>_X)\mul)\lex$.
\end{expl}

%%%%%%%%%%%%%%%%%%%%%%%%%%%%%%%%%%%%%%%%%%%%%%%%%%%%%%%%%%%%%%%%%%%%%%%%%%%%%%
% accessibility
%%%%%%%%%%%%%%%%%%%%%%%%%%%%%%%%%%%%%%%%%%%%%%%%%%%%%%%%%%%%%%%%%%%%%%%%%%%%%%

For building the computability closure, one must start from the
variables of the left hand-side. However, one cannot take any variable
since not every subterm of a computable term is computable {\em a
priori}. To this end, based on the definition of the interpretation of
constant predicate symbols, we introduce the notion of accessibility.

\begin{dfn}[Accessibility]
\label{def-acc}
We say that $u:U$ is {\em $a$-accessible}\footnote{We may not indicate
$a$ if it is not relevant.} in $t:T$, written $t:T\tgt_a u:U$, iff
$t=f\vu$, $f\in\co$, $f:(\vy:\vU)C^{s\alpha}\vv$, $|\vu|=|\vy|$,
$u=u_j$, $j\in\acc(f)$, $T=C^{s\alpha\vphi}\vv\g$, $U= U_j\g\vphi$,
$\g=\vyu$, $\vphi=\{\alpha\to a\}$ and $\pos(\alpha,\vu)=\vide$.

A constructor $c:(\vy:\vU)C^a\vv$ is {\em finitely
branching}\footnote{Primitive types are finitely branching.} iff, for
all $j\in\acc(c)$, either $\pos(\alpha,U_j)=\vide$ or there exists $D$
such that $U_j=D^\alpha\vu$. We say that $u:U$ is {\em strongly
$a$-accessible} in $t:T$, written $t:T\dgd_a~ u:U$, iff $t:T\tgt_a
u:U$, $f$ is a finitely branching constructor and
$\pos(\alpha,U_j)\neq\vide$.

We say that $u:U$ is {\em $*$-accessible modulo $\vphi$} in $t:T$,
written $t:T\gg_\vphi u:U$, iff either $t:T\vphi=u:U$ and
$\vphi|_{\cV(T)}$ is a renaming, or $t:T\vphi\dgd^*\tgt_\ep u:U$ for
some size variable $\ep$.
\end{dfn}

%%%%%%%%%%%%%%%%%%%%%%%%%%%%%%%%%%%%%%%%%%%%%%%%%%%%%%%%%%%%%%%%%%%%%%%%%%%%%%
% termination criterion
%%%%%%%%%%%%%%%%%%%%%%%%%%%%%%%%%%%%%%%%%%%%%%%%%%%%%%%%%%%%%%%%%%%%%%%%%%%%%%

\begin{dfn}[Termination criterion]
  Let $(f\vl\a r,\G,\vphi)\in\cR$ with $f:{(\vx:\vT)}U$ and $\g=\vxl$.
  The {\em computability closure} associated to this rule is given by
  the type system of Figure \ref{fig-thc} on the set of terms
  $\cT_\cA(\cF',\cX')$ where $\cF'=\cF\cup\dom(\G)$,
  $\cX'=\cX\moins\dom(\G)$ and, for all $x\in\dom(\G)$, $\tau_x=x\G$
  and $x<_\cF f$. The termination conditions are:

\begin{lst}{\bu}
\item Well-typedness: for all $x\in\dom(\G)$, $\thc l_i:T_i\vphi\g$.
\item Linearity: $\G$ is linear w.r.t. size variables.
\item Accessibility: for all $x\in\dom(\G)$, there are $i$ and $\b$
such that $l_i:T_i\g\gg_\vphi x:x\G$,\footnote{This implies in
particular that every $x\G$ is of the form $C^\ep\vt$ with
$\ep\in\cZ$.} $T_i=C^\b\vt$ and $\cV(\vt)=\vide$.
\item Computability closure: $\thc r:U\vphi\g$.
\item Positivity: for all $\alpha\in\cV(\vT)$, $\pos(\alpha,U)\sle\pp(U)$.
\item Safeness: $\g$ is an injection from $\domB(\G_f)$ to $\domB(\G)$.
\end{lst}
\end{dfn}

%%%%%%%%%%%%%%%%%%%%%%%%%%%%%%%%%%%%%%%%%%%%%%%%%%%%%%%%%%%%%%%%%%%%%%%%%%%%%%
% rules for the computability closure
%%%%%%%%%%%%%%%%%%%%%%%%%%%%%%%%%%%%%%%%%%%%%%%%%%%%%%%%%%%%%%%%%%%%%%%%%%%%%%

\begin{figure}[ht]
\centering
\caption{Computability closure of $f\vl\a r$ with
$f:(\vx:\vT)U$ and $\g=\vxl$\label{fig-thc}}

\begin{tabular}{rcc}
\\ (ax) & $\cfrac{}{\thc\st:\B}$\\

\\ (size) & $\cfrac{\thc\tC:\B}{\thc C^a:\tC}$ & $(C\in\CFB)$\\

\\ (symb) & $\cfrac{
\thc\tg:s_g \quad (\all i)\D\thc y_i\d:U_i\psi\d}
{\D\thc g\vy\d:V\psi\d}$ &
$\begin{array}{c}
(g\notin\CFB,\, g:(\vy:\vU)V,\\
(g,\psi)<^\cA (f,\vphi))\\
\end{array}$\\

\\ (var) & $\cfrac{\D\thc T:s_x}{\D,x:T\thc x:T}$
& $(x\notin\dom(\D))$\\

\\ (weak) & $\cfrac{\D\thc t:T \quad \D\thc U:s_x}{\D,x:U\thc t:T}$
& $(x\notin\dom(\D))$\\

\\ (prod) & $\cfrac{\D,x:U\thc V:s}{\D\thc (x:U)V:s}$\\

\\ (abs) & $\cfrac{\D,x:U\thc v:V \quad \D\thc (x:U)V:s}
{\D\thc [x:U]v:(x:U)V}$\\

\\ (app) & $\cfrac{\D\thc t:(x:U)V \quad \D\thc u:U}{\D\thc tu:V\xu}$\\

\\ (conv) & $\cfrac{\D\thc t:T \quad \D\thc T:s \quad \D\thc T':s}
{\D\thc t:T'}$ & $(T \le T')$\\
\end{tabular}
\end{figure}

%%%%%%%%%%%%%%%%%%%%%%%%%%%%%%%%%%%%%%%%%%%%%%%%%%%%%%%%%%%%%%%%%%%%%%%%%%%%%%
% comments
%%%%%%%%%%%%%%%%%%%%%%%%%%%%%%%%%%%%%%%%%%%%%%%%%%%%%%%%%%%%%%%%%%%%%%%%%%%%%%

Note that, if $\D\thc t:T$ then $\G,\D\th t:T$. Hence, the
well-typedness condition implies that $\g:\G_f\vphi\leadsto\G$ and
thus that the left hand-side is well-typed: $\G\th f\vl:U\vphi\g$.

The positivity condition on the output type of $f$ w.r.t. size
variables appears in the previous works on sized types too. In
\cite{abel03ita}, Abel gives an example of a function which is not
terminating because it does not satisfy such a condition. This can be
extended to more general continuity conditions
\cite{hughes96popl,abel03tlca} and is indeed necessary (see Example
\ref{expl-mc}).

As for the safeness condition, it simply says that one cannot do
matching or have non-linearities on predicate variables, which is
known to lead to non-termination \cite{harper99ipl}. It is also part
of other works on the Calculus of Constructions with inductive types
\cite{stefanova98thesis} and rewriting \cite{walukiewicz03jfp}.

The positivity, safeness and accessibility conditions are
decidable. For the conditions based on the computability closure, we
prove the strong normalization in Section \ref{sec-prf}.\\

Let us now see some examples.

%% file: expl.tex
%%%%%%%%%%%%%%%%%%%%%%%%%%%%%%%%%%%%%%%%%%%%%%%%%%%%%%%%%%%%%%%%%%%%%%%%%%%%%%
% division
%%%%%%%%%%%%%%%%%%%%%%%%%%%%%%%%%%%%%%%%%%%%%%%%%%%%%%%%%%%%%%%%%%%%%%%%%%%%%%

\begin{expl}[Division on natural numbers, Figure \ref{fig-div}]
Take the types $nat:\st$, $0:nat^0$, $s:nat^\alpha\A nat^{s\alpha}$,
$-:nat^\alpha\A nat^\b\A nat^\alpha$ and $/:nat^\alpha\A nat^\b\A
nat^\alpha$, with $\acc(s)=\{1\}$. All positivity conditions are
clearly satisfied. Safeness is immediate (there is no predicate
variables). For the other conditions, we only detail (3) and (5).

\begin{lst}{\bu}
\item For (3), take $\G_-=p:nat^\alpha,q:nat^\b$, $\z_-(\alpha,\b)=\alpha$,
$\G=x:nat^\d,y:nat^\ep$, $\g=\{p\to sx, q\to sy\}$, $\vphi=\{\alpha\to
s\d,\b\to s\ep\}$ and $s<_\cF -$.

\begin{lst}{--}
\item Well-typedness: By (symb), $\thc x:nat^\d$ and $\thc y:nat^\ep$.
Thus, by (symb), $\thc sx:nat^{s\d}$ and $\thc sy:nat^{s\ep}$.

\item Accessibility: One can easily check that $sx:nat^{s\d}\gg_\vphi
x:nat^\d$ and $sy^{s\ep}\gg_\vphi y:nat^\ep$.

\item Computability closure: By (symb), $\thc x:nat^\d$ and $\thc y:nat^\ep$.
By (symb), $\thc -xy:nat^\d$ since $\z_-(\d,\ep)=\d<
\z_-(s\d,s\ep)=s\d$. Thus, by (sub), $\thc -xy:nat^{s\d}$.
\end{lst}

\item For (5), take $\G_/=p:nat^\alpha,q:nat^\b$, $\z_/(\alpha,\b)=\alpha$,
$\G=x:nat^\d,y:nat^\ep$, $\g=\{p\to sx, q\to y\}$, $\vphi=\{\alpha\to
s\d,\b\to\ep\}$ and $-<_\cF /$.

\begin{lst}{--}
\item Well-typedness: By (symb), $\thc x:nat^\d$ and $\thc y:nat^\ep$.
Thus, by (symb), $\thc sx:nat^{s\d}$.

\item Accessibility: One can easily check that
$sx:nat^{s\d}\gg_\vphi x:nat^\d$ and $y:nat^\ep\gg_\vphi y:nat^\ep$.

\item Computability closure: By (symb), $\thc x:nat^\d$ and $\thc y:nat^\ep$.
By (symb), $\thc -xy:nat^\d$. By (symb), $\thc /(-xy)y:nat^\d$ since
$\z_/(\d,\ep)=\d< \z_/(s\d,\ep)=s\d$. Thus, by (symb), $\thc
s(/(-xy)y): nat^{s\d}$.
\end{lst}
\end{lst}
\end{expl}

%%%%%%%%%%%%%%%%%%%%%%%%%%%%%%%%%%%%%%%%%%%%%%%%%%%%%%%%%%%%%%%%%%%%%%%%%%%%%%
% ordinals
%%%%%%%%%%%%%%%%%%%%%%%%%%%%%%%%%%%%%%%%%%%%%%%%%%%%%%%%%%%%%%%%%%%%%%%%%%%%%%

\begin{expl}[Addition on Brouwer's ordinals, Figure \ref{fig-ord}]
Take the types $ord:\st$, $0:nat^0$, $s:nat^\alpha\A nat^{s\alpha}$,
$lim:(nat\A ord^\alpha)\A ord^{s\alpha}$ and $+:nat^\alpha\A nat^\b\A
nat^\infty$, with $\acc(s)=\acc(lim)=\{1\}$. All positivity conditions
are clearly satisfied. We only detail rule (3). Take
$\G_+=p:ord^\alpha,q:ord^\b$, $\z_+(\alpha,\b)=\alpha$,
$\G=f:nat^\infty\A ord^\d,y:ord^\ep$, $\g=\{p\to limf, q\to y\}$,
$\vphi=\{\alpha\to s\d,\b\to\ep\}$ and $s,lim<_\cF +$.

\begin{lst}{--}
\item Well-typedness: By (symb), $\thc f:nat^\infty\A ord^\d$ and
$\thc y:ord^\ep$. Thus, by (symb), $\thc limf:ord^{s\d}$.
\item Accessibility: One can easily check that $limf:ord^{s\d}\gg_\vphi
f:nat^\infty\A ord^\d$ and $y:ord^\ep\gg_\vphi y:ord^\ep$.
\item Computability closure: By (symb), $\thc
f:nat^\infty\A ord^\d$ and $\thc y:ord^\ep$. Let $\D=x:nat^\infty$. By
(var), $\D\thc x:nat^\infty$. By (weak), $\D\thc f:nat^\infty\A
ord^\d$ and $\D\thc y:ord^\ep$. By (app), $\D\thc fx:ord^\d$. By
(symb), $\D\thc +(fx)y:ord^\infty$ since $\z_+(\d,\ep)=\d<
\z_+(s\d,\ep)=s\d$. By (abs), $\thc [x:nat^\infty](+(fx)y):
(x:nat^\infty)ord^\d$. Thus, by (symb), $\thc
lim([x:nat^\infty](+(fx)y)): ord^{s\d}$.
\end{lst}
\end{expl}

%%%%%%%%%%%%%%%%%%%%%%%%%%%%%%%%%%%%%%%%%%%%%%%%%%%%%%%%%%%%%%%%%%%%%%%%%%%%%%
% rules for quick sort
%%%%%%%%%%%%%%%%%%%%%%%%%%%%%%%%%%%%%%%%%%%%%%%%%%%%%%%%%%%%%%%%%%%%%%%%%%%%%%

\begin{figure}[ht]
\caption{Quick sort\label{fig-qs}}
\begin{center}
$\begin{array}{rrcll}
(1) & \fst~(pair~x~y) &\a& x\\
(2) & snd~(pair~x~y) &\a& y\\[2mm]

(3) & \le~0~x &\a& true\\
(4) & \le~(s~x)~0 &\a& false\\
(5) & \le~(s~x)~(s~y) &\a& \le~x~y\\[2mm]

(6) & \si~true~x~y &\a& x\\
(7) & \si~false~x~y &\a& y\\[2mm]

(8) & pivot~x~nil &\a& pair~nil~nil\\
(9) & pivot~x~(cons~y~l) &\a&
\si~(\le~y~x)~(pair~(cons~y~u)~v)~(pair~u~(cons~y~v))\\
&&& \mbox{where } u=\fst~(pivot~x~l) \mbox{ and } v=snd~(pivot~x~l)\\[2mm]

(10) & qs~nil~l &\a& l\\
(11) & qs~(cons~x~l)~l' &\a& qs~u~(cons~x~(qs~v~l'))\\
&&& \mbox{where } u=\fst~(pivot~x~l) \mbox{ and } v=snd~(pivot~x~l)\\[2mm]

(12) & qsort~l &\a& qs~l~nil\\
\end{array}$
\end{center}
\end{figure}

%%%%%%%%%%%%%%%%%%%%%%%%%%%%%%%%%%%%%%%%%%%%%%%%%%%%%%%%%%%%%%%%%%%%%%%%%%%%%%
% quick sort
%%%%%%%%%%%%%%%%%%%%%%%%%%%%%%%%%%%%%%%%%%%%%%%%%%%%%%%%%%%%%%%%%%%%%%%%%%%%%%

\begin{expl}[Quick sort, Figure \ref{fig-qs}]
\label{expl-qs}
Take the types $bool:\st$, $true:bool^\infty$, $false:bool^\infty$,
$list:\st$, $nil:list^0$, $cons:nat^\infty\A list^\alpha\A
list^{s\alpha}$, $blist:\st$, $pair:list^\alpha\A list^\b\A
blist^{max(\alpha,\b)}$, $\fst:blist^\alpha\A list^\alpha$,
$snd:blist^\alpha\A list^\alpha$, $\le:nat^\infty\A nat^\infty\A
bool^\infty$, $pivot:nat^\infty\A list^\alpha\A blist^\alpha$,
$qs:list^\infty\A list^\infty\A list^\infty$ and $qsort:list^\infty\A
list^\infty$. We only detail the computability closure condition of
rule (11).

Take $\z_{qs}(\alpha,\b)=\alpha$,
$\G=x:nat^\infty,l:list^\d,l':list^\ep$, $\vphi=\{\alpha\to
s\d,\b\to\ep\}$ and $qs >_\cF pivot >_\cF cons, pair, \fst, snd$. By
(symb), $\thc x:nat^\infty$, $\thc l:list^\d$ and $\thc
l':list^\ep$. By (symb), $\thc pivot~x~l:blist^\d$. By (symb), $\thc
u:list^\d$ and $\thc v:list^\d$. By (symb), $\thc
qs~v~l':list^\infty$. By (symb), $\thc
cons~x~(qs~v~l'):list^\infty$. Thus, by (symb), $\thc
qs~u~(cons~x~(qs~v~l')):list^\infty$ since $\z_{qs}(\d,\infty)=\d<
\z_{qs}(s\d,\ep)=s\d$.

Note that we cannot take $qs:list^\alpha\A list^\b\A list^{\alpha+\b}$
and thus $qsort:list^\alpha\A list^\alpha$ since too much information
is lost by taking $pair:list^\alpha\A list^\b\A
blist^{max(\alpha,\b)}$. Even though we take $pair:list^\alpha\A
list^\b\A blist^{\langle\alpha,\b\rangle}$ with
$\langle\alpha,\b\rangle$ interpreted as a pair of ordinals, the
current setting does not allow us to say that $pivot$ has type
$nat^\infty\A list^\alpha\A blist^{\langle\b,\g\rangle}$ for some $\b$
and $\g$ such that $\b+\g=\alpha$, as it can be done in Xi's framework
\cite{xi02hosc}.
\end{expl}

The following examples are taken from \cite{giesl97jar}.

%%%%%%%%%%%%%%%%%%%%%%%%%%%%%%%%%%%%%%%%%%%%%%%%%%%%%%%%%%%%%%%%%%%%%%%%%%%%%%
% rules for if
%%%%%%%%%%%%%%%%%%%%%%%%%%%%%%%%%%%%%%%%%%%%%%%%%%%%%%%%%%%%%%%%%%%%%%%%%%%%%%

\begin{figure}[ht]
\caption{Paulson's normalization of $\si$-expressions\label{fig-if}}
\begin{trans}
(1) & nm~at & at\\
(2) & nm~(\si~at~y~z) & \si~at~(nm~y)~(nm~z)\\
(3) & nm~(\si~(\si~u~v~w)~y~z) & nm~(\si~u~(nm~(\si~v~y~z))~(nm~(\si~w~y~z)))\\
\end{trans}
\end{figure}

%%%%%%%%%%%%%%%%%%%%%%%%%%%%%%%%%%%%%%%%%%%%%%%%%%%%%%%%%%%%%%%%%%%%%%%%%%%%%%
% if
%%%%%%%%%%%%%%%%%%%%%%%%%%%%%%%%%%%%%%%%%%%%%%%%%%%%%%%%%%%%%%%%%%%%%%%%%%%%%%

\begin{expl}[Paulson's normalization of $\si$-expressions, Figure \ref{fig-if}]
\label{expl-if}
Take the types $expr:\st$, $at:expr^1$, $\si:expr^\alpha\A expr^\b\A
expr^\g\A expr^{\alpha(1+\b+\g)}$ and $nm:expr^\alpha\A
expr^\alpha$. We only detail the computability closure condition of
rule (3). Take $\z_{nm}(\alpha)=\alpha$, $\G=u:expr^\alpha, v:expr^\b,
w:expr^\g, y:expr^\d, z:expr^\ep$, $\up=\alpha(1+\b+\g)(1+\d+\ep)$,
$\vphi=\{\alpha\to\up\}$ and $nm>_\cF at,\si$. Then, one can check that
$\up$ is strictly greater than $\b(1+\d+\ep)$, $\g(1+\d+\ep)$ and
$\alpha(1+\b(1+\d+\ep)+\g(1+\d+\ep))$ if variables are interpreted by
strictly positive integers.
\end{expl}

%%%%%%%%%%%%%%%%%%%%%%%%%%%%%%%%%%%%%%%%%%%%%%%%%%%%%%%%%%%%%%%%%%%%%%%%%%%%%%
% rules for reverse
%%%%%%%%%%%%%%%%%%%%%%%%%%%%%%%%%%%%%%%%%%%%%%%%%%%%%%%%%%%%%%%%%%%%%%%%%%%%%%

\begin{figure}[ht]
\caption{Huet and Hullot's reverse function\label{fig-rev}}
\begin{trans}
(1) & rev1~x~nil & x\\
(2) & rev1~x~(cons~y~l) & rev1~y~l\\[2mm]

(3) & rev2~x~nil & nil\\
(4) & rev2~x~(cons~y~l) & rev~(cons~x~(rev~(rev2~y~l)))\\[2mm]

(5) & rev~nil & nil\\
(6) & rev~(cons~x~l) & cons~(rev1~x~l)~(rev2~x~l)\\
\end{trans}
\end{figure}

%%%%%%%%%%%%%%%%%%%%%%%%%%%%%%%%%%%%%%%%%%%%%%%%%%%%%%%%%%%%%%%%%%%%%%%%%%%%%%
% reverse
%%%%%%%%%%%%%%%%%%%%%%%%%%%%%%%%%%%%%%%%%%%%%%%%%%%%%%%%%%%%%%%%%%%%%%%%%%%%%%

\begin{expl}[Huet and Hullot's reverse function, Figure \ref{fig-rev}]
\label{expl-rev}
Take the types $rev1:nat^\infty\A list^\infty\A nat^\infty$,
$rev2:nat^\infty\A list^\beta\A list^\beta$ and $rev:list^\alpha\A
list^\alpha$. We only detail the computability closure condition of
rule (4). Take $\z_{rev}(\alpha)=2\alpha$,
$\z_{rev2}(\alpha,\b)=2\b+1$,
$\G=x:nat^\infty,y:nat^\infty,l:list^\d$, $\vphi=\{\beta\to \d+1\}$
and $rev\simeq_\cF rev2>_\cF rev1>_\cF cons,nil$. Then, one can check
that $\z_{rev2}(\infty,\d+1)=2\d+3$ is strictly greater than
$\z_{rev2}(\infty,\d)=2\d+1$, $\z_{rev}(\d)=2\d$ and
$\z_{rev}(1+\d)=2\d+2$.
\end{expl}

%%%%%%%%%%%%%%%%%%%%%%%%%%%%%%%%%%%%%%%%%%%%%%%%%%%%%%%%%%%%%%%%%%%%%%%%%%%%%%
% rules for mc
%%%%%%%%%%%%%%%%%%%%%%%%%%%%%%%%%%%%%%%%%%%%%%%%%%%%%%%%%%%%%%%%%%%%%%%%%%%%%%

\begin{figure}[ht]
\caption{Mac Carthy's ``91'' function\label{fig-mc}}
\begin{center}
$\begin{array}{rr@{~~\a~~}l@{~~\mbox{if}~~}l}
(1) & f~x & f~(f~(+~x~11)) & \le~x~100 = true\\
(2) & f~x & -~x~10 & \le~x~100 = false\\
\end{array}$
\end{center}
\end{figure}

%%%%%%%%%%%%%%%%%%%%%%%%%%%%%%%%%%%%%%%%%%%%%%%%%%%%%%%%%%%%%%%%%%%%%%%%%%%%%%
% mac carthy
%%%%%%%%%%%%%%%%%%%%%%%%%%%%%%%%%%%%%%%%%%%%%%%%%%%%%%%%%%%%%%%%%%%%%%%%%%%%%%

\begin{expl}[Mac Carthy's ``91'' function, Figure \ref{fig-mc}]
\label{expl-mc}
\hfill Mac Carthy's ``91''\\ function $f$ is defined by the following
equations: $f(x)= f(f(x+11))$ if $x\le 100$, and $f(x)=x-10$
otherwise. In fact, one can prove that $f$ is equal to the function
$F$ such that $F(x)=91$ if $x\le 100$, and $F(x)=x-10$ otherwise. A
way to formalize this in CACSA would be to use conditional rewrite
rules (see Figure \ref{fig-mc}) and take\footnote{Note that
$F(\alpha)$ is monotone w.r.t. $\alpha$.} $f:nat^\alpha\A
nat^{F(\alpha)}$ and $\z_f^X(x)=max(0,101-x)$ as measure function, as
it can be done in Xi's framework. Then, by taking into account the
rewrite rule conditions, one could prove that, if $\G=x:nat^\d$ and
$\le~x~100=true$, then $\d\le 100$, $\z_f(\d+11)<\z_f(\d)$ and
$\z_f(F(\d))<\z_f(\d)$.
\end{expl}

%% file: prf.tex
%%%%%%%%%%%%%%%%%%%%%%%%%%%%%%%%%%%%%%%%%%%%%%%%%%%%%%%%%%%%%%%%%%%%%%%%%%%%%%
% termination proof
%%%%%%%%%%%%%%%%%%%%%%%%%%%%%%%%%%%%%%%%%%%%%%%%%%%%%%%%%%%%%%%%%%%%%%%%%%%%%%

\section{Termination proof}
\label{sec-prf}

We first prove some lemmas for proving the correctness of
accessibility w.r.t. computability (accessible subterms of a
computable term are computable). Then, we prove the correctness of the
computability closure (every term of the computability closure is
computable) and the computability of every symbol, hence the strong
normalization of every well-typed term.

%%%%%%%%%%%%%%%%%%%%%%%%%%%%%%%%%%%%%%%%%%%%%%%%%%%%%%%%%%%%%%%%%%%%%%%%%%%%%%
% lemmas on accessibility
%%%%%%%%%%%%%%%%%%%%%%%%%%%%%%%%%%%%%%%%%%%%%%%%%%%%%%%%%%%%%%%%%%%%%%%%%%%%%%

\begin{lem}[Accessibility properties]
\label{lem-acc}
\hfill
\begin{enumi}{}
\item If $t:T\dgd^k~ u:D^e\vu$ then $T=C^{s^ke}\vt$.
\item If $t:C^\b\vt\gg_\vphi u:U$ then there are $\ep\in\cZ$ and $k\ge 0$
such that $\b\vphi=s^k\ep$.
\item If $t:T\tgt u:U$, $t\s\in K_C^\fb(\vS)$ then
$o_{C(\vS)}(t)$ is a successor ordinal.
\item If $t:T\dgd~ u:U$ and $t\s\in I_C^\fb(\vS)$ then $u\s\in I_D^\fb(\vS')$
for some $D$ and $\vS'$.
\item Let $f:(\vy:\vU)C^{s\alpha}\vv$ be a finitely branching constructor
such that, if $j\in\acc(f)$ and $\pos(\alpha,U_j)\neq\vide$ then
$U_j=C_j^\alpha\vv^j$. If $f\vu\in K_C^\fa(\vS)$ then
$o_{C(\vS)}(f\vu)= max \{o_{C_j(\vS^j)}(u_j)~|~ j\in\acc(f),
\pos(\alpha,U_j)\neq\vide\} + 1$, where
$\vS^j=\I{\vv^j}_{\xi,\t}^\nu$, $y\xi=S_{\io_y}$, $\vy\t=\vu$ and
$\alpha\nu=\fa$.
\item If $t:T\dgd^k\tgt~ u:U$ and $t\s\in K_C^\fb(\vS)$ then
$o_{C(\vS)}(t)=\fa+k+1$ for some $\fa$.
\item If $t:T\tgt^* u:U$ and $t\s\in \I{T}^\mu_{\xi,\s}$ then $u\s\in
\I{U}^\mu_{\xi,\s}$.
\end{enumi}
\end{lem}

\begin{prf}
\begin{enumi}{}
\item By induction on $k$. For $k=0$, this is immediate. Assume now
that $t:T\dgd^k~ v:V\dgd_a~ u:D^e\vu$. Then, $a=e$ and
$V=E^{se}\vv\g$. Therefore, by induction hypothesis,
$T=C^{s^{k+1}e}\vt$.

\item There are two cases.
\begin{lst}{--}
\item $t:C^\b\vphi=u:U$ and $\vphi_{|\cV(T)}$ is a renaming.
Take $\ep=\b\vphi$ and $k=0$.
\item $t:C^\b\vphi\dgd^k~ v:V\tgt_\ep u:U$.
Then, $V=D^{s\ep}\vv$ and, by (1), $\b\vphi=s^{k+1}\ep$.
\end{lst}

\item By Lemma \ref{lem-suc}.

\item By (3), we can assume that $t\s\in I_C^{\fa+1}(\vS)$. By Definition
\ref{def-int-const}, $u_j\in \I{U_j}_{\xi,\t}^\nu$ with $y\xi=S_{\io_y}$,
$\vy\t=\vu$ and $\alpha\nu=\fa$. By definition of $\!\!~\dgd$,
$U_j=D^\alpha\vu$. Thus, $u_j\in I_D^\fa(\vS')$ with
$\vS'=\I{\vu}_{\xi,\t}^\nu$.

\item By (3), we can assume that $f\vu\in I_C^{\fa+1}(\vS)$. By (4),
for all $j\in\acc(f)$ such that $\pos(\alpha,U_j)\neq\vide$, $u_j\in
I_{C_j}^\fa(\vS^j)$. Let $\fa_j=o_{C_j(\vS^j)}(u_j)$. Since $\fa$ is
as small as possible, we must have $max
\{\fa_j~|~j\in\acc(f),\pos(\alpha,U_j)\neq\vide\}=\fa$.

\item By induction on $k$. For $k=0$, this is (3). Assume now that
$t:T\dgd~ u:U\dgd^k\tgt~ v:V$. By (4), for all $j\in\acc(f)$,
$u_j\s\in I_{D_j}^\fa(\vS^j)$. Let $\fa_j=o_{C_j(\vS^j)}(u_j\s)$. By
induction hypothesis, $\fa_j=\fb_j+k+1$. Therefore, by (5),
$o_{C(\vS)}(t\s)=\fb_j+k+2$ for some $\fb_j$.

\item By induction on the number of $\tgt$-steps. If there is no step, this
is immediate. Assume now that $t:T\tgt_a u:U\tgt^* v:V$ and
$\alpha\vphi=a$. Since $T=C^{s\alpha\vphi}\vv\g$, $\I{T}^\mu_{\xi,\s}=
I_C^{\alpha\vphi\mu+1}(\vS)$ with
$\vS=\I{\vv\g}^\mu_{\xi,\s}$. Therefore, $u\s\in
\I{U_j}^{\vphi\mu}_{\eta,\g\s}$ with $y\eta=S_{\io_y}$. Since
$v_{\io_y}=y$, $y\eta= \I{y\g}^{\vphi\mu}_{\xi,\s}=
\I{y\g}^\mu_{\xi,\s}$ since $\pos(\alpha,\g)=\vide$. So, by candidate
substitution, $\I{U_j}^{\vphi\mu}_{\eta,\g\s}=
\I{U_j\g}^{\vphi\mu}_{\xi,\s}= \I{U}^\mu_{\xi,\s}$. Therefore, by induction
hypothesis, $v\s\in \I{V}^\mu_{\xi,\s}$.\cqfd
\end{enumi}
\end{prf}

%%%%%%%%%%%%%%%%%%%%%%%%%%%%%%%%%%%%%%%%%%%%%%%%%%%%%%%%%%%%%%%%%%%%%%%%%%%%%%
% correctness of accessibility
%%%%%%%%%%%%%%%%%%%%%%%%%%%%%%%%%%%%%%%%%%%%%%%%%%%%%%%%%%%%%%%%%%%%%%%%%%%%%%

\begin{thm}[Accessibility correctness]
\label{thm-acc}
If $t:T\gg_\vphi u:U$, $T=C^\b\vt$, $\cV(\vt)=\vide$ and $t\s\in
\I{T}^\mu_{\xi,\s}$ then there exists $\nu$ such that
$\b\vphi\nu\le\b\mu$ and $u\s\in \I{U}^\nu_{\xi,\s}$.
\end{thm}

\begin{prf}
There are two cases:

\begin{lst}{\bu}
\item $t:T\vphi=u:U$ and $\vphi_{|\cV(T)}$ is a renaming. Let
$\nu=\vphi_{|\cV(T)}^{-1}\mu$. $\b\vphi\nu=\b\mu$ and $u\s=t\s\in
\I{T}_{\xi,\s}^\mu= \I{T\vphi}_{\xi,\s}^\nu$.

\item $t:T\vphi\dgd^* u:U\tgt_\ep v:V$. By definition of $\tgt_\ep$,
$U=D^{s\ep}\vu$. By Lemma \ref{lem-acc} (1), $\b\vphi=s^{k+1}\ep$. By
(6), there exists $\fa$ such that $\fa+k+1\le\b\mu$ and $t\s\in
I_C^{\fa+k+1}(\vS)$. Let $\ep\nu=\fa$. Then,
$\b\vphi\nu=s^{k+1}\ep\nu=\fa+k+1\le\b\mu$, $t\s\in
\I{T\vphi}^\nu_{\xi,\s}$ and, by (7), $u\s\in
\I{T\vphi}^\nu_{\xi,\s}$.\cqfd
\end{lst}
\end{prf}

%%%%%%%%%%%%%%%%%%%%%%%%%%%%%%%%%%%%%%%%%%%%%%%%%%%%%%%%%%%%%%%%%%%%%%%%%%%%%%
% correctness of computability closure
%%%%%%%%%%%%%%%%%%%%%%%%%%%%%%%%%%%%%%%%%%%%%%%%%%%%%%%%%%%%%%%%%%%%%%%%%%%%%%

\begin{thm}[Correctness of the computability closure]
\label{thm-cor-thc}

Let $(f\vl\a r,\G$, $\vphi)\in\cR$, $f:(\vx:\vT)U$ and
$\g=\vxl$. Assume that, for all $(g,\mu)<^\fA(f,\vphi\nu)$,
$g\in\I\tg^\mu$. If $\D\thc t:T$ and $\xi,\s\models_\nu\G,\D$ then
$t\s\in \I{T}^\nu_{\xi,\s}$.
\end{thm}

\begin{prf}
  By induction on $\D\thc t:T$. We only detail the case (symb). Since
  $(g,\psi)<^\cA(f,\vphi)$, $(g,\psi\nu)<^\fA(f,\vphi\nu)$. Hence, by
  assumption, $g\in\I\tg^{\psi\nu}$. Now, by induction hypothesis,
  $\vy\d\s\in \I{\vU\psi\d}^\nu_{\xi,\s}$. By candidate substitution,
  there exists $\eta$ such that $\I{\vU\psi\d}^\nu_{\xi,\s}=
  \I{\vU\psi}^\nu_{\eta,\d\s}$. By size substitution,
  $\I{\vU\psi}^\nu_{\eta,\d\s}=
  \I\vU^{\psi\nu}_{\eta,\d\s}$. Therefore, $g\vy\d\s\in
  \I{V}^{\psi\nu}_{\eta,\d\s}= \I{V\psi\d}^\nu_{\xi,\s}$.
\end{prf}

%%%%%%%%%%%%%%%%%%%%%%%%%%%%%%%%%%%%%%%%%%%%%%%%%%%%%%%%%%%%%%%%%%%%%%%%%%%%%%
% computability of symbols
%%%%%%%%%%%%%%%%%%%%%%%%%%%%%%%%%%%%%%%%%%%%%%%%%%%%%%%%%%%%%%%%%%%%%%%%%%%%%%

\begin{lem}[Computability of symbols]
\label{lem-comp-symb}
For all $f$ and $\mu$, $f\in\I\tf^\mu$.
\end{lem}

\begin{prf}
  Assume that $\tf=(\vx:\vT)U$ with $U$ distinct from a
  product. $f\in\I\tf^\mu$ iff, for all $\eta,\t$ such that
  $\eta,\t\models_\mu\G_f$, $f\vx\t\in \I{U}^\mu_{\eta,\t}$. We prove
  it by induction on $((f,\mu),\t)$ with ${(>^\fA,\a)\lex}$ as
  well-founded ordering. Let $t_i= x_i\t$ and $t=f\vt$. By assumption,
  for every rule $f\vl\a r\in\cR$, $|\vl|\le|\vt|$. So, if $f\notin
  Cons$ then $t$ is neutral and it suffices to prove that
  $\a\!\!(t)\sle \I{U}^\mu_{\eta,\t}$. Otherwise,
  $\I{U}^\mu_{\eta,\t}= I_C^{a\mu}(\vS)$ with
  $\vS=\I\vv^\mu_{\eta,\t}$. Since $\eta,\t\models_\mu\G_f$, $t_j\in
  \I{T_j}^\mu_{\eta,\t}$. Therefore, in this case too, it suffices to
  prove that $\a\!\!(t)\sle \I{U}^\mu_{\eta,\t}$.
  
  If the reduction takes place in one $t_i$ then we can conclude by
  induction hypothesis. Assume now that there exist $(l\a
  r,\G,\vphi)\in\cR$ and $\s$ such that $t=l\s$. Then, $l=f\vl$ and
  $\t=\g\s$ with $\g=\vxl$.

  We now define $\xi$ such that $\I{U}^\mu_{\eta,\g\s}=
  \I{U\g}^\mu_{\xi,\s}$ and $\I\vT^\mu_{\eta,\g\s}=
  \I{\vT\g}^\mu_{\xi,\s}$. By safeness, $\g$ is an injection from
  $\domB(\G_f)$ to $\domB(\G)$. Let $y\in\domB(\G)$. If there exists
  $x\in\dom(\G_f)$ (necessarily unique) such that $y=x\g$, we take
  $y\xi=x\eta$. Otherwise, we take $y\xi= \top_{y\G}$.

  We check that $\xi\models\G$. If $y\neq x\g$, $y\xi= \top_{y\G}\in
  \cR_{y\G}$. If $y=x\g$ then $y\xi=x\eta$. Since $\eta\models\G_f$,
  $x\eta\in \cR_{x\G_f}$. Since $\g:\G_f\vphi\leadsto\G$, $\G\th
  y:x\G_f\vphi\g$. Therefore, $y\G\le x\G_f\vphi\g$ and $\cR_{y\G}=
  \cR_{x\G_f\vphi\g}= \cR_{x\G_f}$. So, $\xi\models\G$.

  Now, by candidate substitution, $\I{U\g}^\mu_{\xi,\s}=
  \I{U}^\mu_{\eta',\g\s}$ with $x\eta'= \I{x\g}_{\xi,\s}$. Let
  $x\in\FV(\vT U)$. By safeness, $x\g= y\in \domB(\G)$ and $x\eta'=
  y\xi= x\eta$. Therefore, $\eta'=\eta$.

  We now prove that $\xi,\s\models_\nu\G$ for some valuation $\nu$
  such that $\vphi\nu\le\mu$. Let $x\in\dom(\G)$. By assumption, there
  exists $i$ such that $l_i:T_i\g\gg_\vphi x:x\G$, $T_i\g=C^{\b_x}\vu$
  and $\cV(\vu)=\vide$. By Lemma \ref{lem-acc} (2), there is $\ep_x$
  and $k_x$ such that $\b_x\vphi=s^{k_x}\ep_x$. Since $l_i\s\in
  \I{T_i\g}_{\xi,\s}$, by Theorem \ref{thm-acc}, there exists $\nu_x$
  such that $x\s\in \I{x\G}^{\nu_x}_{\xi,\s}$ and
  $\b_x\vphi\nu_x\le\b_x\mu$. Since $\G$ is linear w.r.t. size
  variables, $\ep_x\neq\ep_y$ whenever $x\neq y$. So, we can define
  $\nu$ by taking $\ep_x\nu= \ep_x\nu_x$. Then, $\b_x\vphi\nu=
  s^{k_x}\ep_x\nu= s^{k_x}\ep_x\nu_x= \b_x\vphi\nu_x\le \b_x\mu$.

  Therefore, since $\thc r:U\vphi\g$, by correctness of the
  computability closure, $r\s\in \I{U\vphi\g}^\nu_{\xi,\s}=
  \I{U\vphi}^\nu_{\eta,\t}= \I{U}^{\vphi\nu}_{\eta,\t}\le
  \I{U}^\mu_{\eta,\t}$ since, for all $\alpha$,
  $\pos(\alpha,U)\sle\pos^+(U)$.\cqfd
\end{prf}

%%%%%%%%%%%%%%%%%%%%%%%%%%%%%%%%%%%%%%%%%%%%%%%%%%%%%%%%%%%%%%%%%%%%%%%%%%%%%%
% strong normalization
%%%%%%%%%%%%%%%%%%%%%%%%%%%%%%%%%%%%%%%%%%%%%%%%%%%%%%%%%%%%%%%%%%%%%%%%%%%%%%

\begin{thm}[Strong normalization]
\label{thm-sn}
Every well-typed term is strongly normalizable.
\end{thm}

\begin{prf}
The invariance by reduction is proved in \cite{blanqui03mscs}. Hence,
we can conclude by Theorem \ref{thm-cond-sn} and Lemma
\ref{lem-comp-symb}.\cqfd
\end{prf}

%% file: conclu.tex
%%%%%%%%%%%%%%%%%%%%%%%%%%%%%%%%%%%%%%%%%%%%%%%%%%%%%%%%%%%%%%%%%%%%%%%%%%%%%%
% conclusion
%%%%%%%%%%%%%%%%%%%%%%%%%%%%%%%%%%%%%%%%%%%%%%%%%%%%%%%%%%%%%%%%%%%%%%%%%%%%%%

\section{Conclusion}

The notion of computability closure, first introduced in
\cite{blanqui02tcs} and further extended to higher-order
pattern-matching \cite{blanqui00rta}, higher-order recursive path
ordering \cite{jouannaud99lics}, type-level rewriting
\cite{blanqui01lics} and rewriting modulo equational theories
\cite{blanqui03rta}, again shows to be essential for extending to
rewriting and dependent types type-based termination criteria for
(polymorphic) $\la$-calculi with inductive types and case analysis
\cite{hughes96popl,xi02hosc,barthe04mscs,abel02tr}. In contrast with
what is suggested in \cite{barthe04mscs}, this notion, which is
expressed as a sub-system of the whole type system (by restricting the
size of arguments in function calls in some computability-preserving
way, see Figure \ref{fig-thc}), allows pattern-matching and does not
suffer from limitations one could find in systems relying on external
guard predicates for recursive definitions.

Moreover, we allow a richer size algebra than the one in
\cite{hughes96popl,barthe04mscs,abel02tr} (see Section
\ref{sec-cons}). But, we do not allow existential size variables and
conditional rewriting that are essential for capturing for instance
the size-preserving property of quicksort (Example \ref{expl-qs}) and
Mac Carty's ``91'' function (Example \ref{expl-mc}) respectively, as
it can be done in Xi's work \cite{xi02hosc}. Such extensions should
allow us to subsume Xi's work completely. More generally, it is
important to have a better understanding of the differences between
Xi's work which does not use subtyping (but has existential size
variables and singleton types) and the other works that are based on
subtyping.

In this work, we assume that users provide appropriate sized types for
function symbols and then check by our technique that the rewrite
rules defining these function symbols are compatible with their
types. An important extension would be to infer these types. Works in
this direction for ML-like languages are
\cite{nelson95thesis,zenger97tcs,chin01hosc}. The exact relations between
these works and with refinement types also
\cite{pfenning93types,freeman94thesis} still have to be investigated.
Note also that deciding the non-size-increasing property of some
functions is investigated in \cite{giesl95ki,giesl95sas}.

We made two important assumptions that also need further
research. First, the confluence of $\b\cup\cR$, which is still an open
problem when $\cR$ is confluent, terminating, non left-linear and
contains type-level rewrite rules. Second, the preservation of typing
under rewriting (subject reduction for $\cR$), for which we need to
find decidable sufficient conditions (see Example \ref{expl-sr}).

Finally, by combining rewriting and subtyping in the Calculus of
Constructions, this work may also be seen as an important step towards
the integration of membership equational logic \cite{bouhoula00tcs}
and dependent type systems. Previous works in this direction are
\cite{barthe00fossacs,castagna01ic,stehr02thesis}.\\

{\bf Acknowledgments.} I would like to thank very much Ralph Matthes
for having invited me for a one-week stay in M\"unich in February
2002. Andreas Abel's technical report \cite{abel02tr} and the
discussions I had with Ralph and Andreas about monotone inductive
types and termination were the starting point of the present work.

%% file: trans.tex
%%%%%%%%%%%%%%%%%%%%%%%%%%%%%%%%%%%%%%%%%%%%%%%%%%%%%%%%%%%%%%%%%%%%%%%%%%%%%%
% admissibility of transitivity
%%%%%%%%%%%%%%%%%%%%%%%%%%%%%%%%%%%%%%%%%%%%%%%%%%%%%%%%%%%%%%%%%%%%%%%%%%%%%%

\section{Elimination of transitivity}
\label{sec-trans}

In this section, we prove Theorem \ref{thm-trans-elim} by following
Chen's technique \cite{chen98thesis}.

\begin{lem}
$\le$ is equivalent to the relation $\le'$ where (symb) is
replaced by:

\begin{center}
(symb')\quad $\cfrac{C^b\vt\le T}{C^a\vt\le T}$\quad $(a\le_\cA b)$
\end{center}
\end{lem}

\begin{prf}
$\le\sle\le'$: Assume that $a\le_\cA b$. By (refl), $C^b\vt\le'
C^b\vt$. Hence, by (symb'), $C^a\vt\le' C^b\vt$. $\le'\sle\le$: Assume
that $C^a\vt\le' T$ since $C^b\vt\le' T$ and $a\le_\cA b$. By
induction hypothesis, $C^b\vt\le T$. By (symb), $C^a\vt\le
C^b\vt$. Therefore, by (trans), $C^a\vt\le T$.\cqfd\\
\end{prf}

Note that the following two subtyping rules are clearly admissible:

\begin{center}
\begin{tabular}{rc}
(left)\quad $\cfrac{T\ad T' \quad T'\le U}{T\le U}$\\
\\(right)\quad $\cfrac{T\le U' \quad U'\ad U}{T\le U}$\\
\end{tabular}
\end{center}

For representing the subtyping deductions, we introduce the following
term algebra:

\begin{center}
$d ::= \bot ~|~ I ~|~ Sd ~|~ Cd ~|~ Ld ~|~ Rd ~|~ Pdd ~|~ Tdd$
\end{center}

\noindent
where $\bot$ stands for some impossible case, $I$ for (refl), $S$ for
(symb'), $C$ for (conv), $L$ for (left), $R$ for (right), $P$ for
(prod), and $T$ for (trans).

We now prove that the transformation rules of Figure \ref{fig-trans}
are valid, that is, a deduction matching a left hand-side can be
replaced by the corresponding right hand-side.

%%%%%%%%%%%%%%%%%%%%%%%%%%%%%%%%%%%%%%%%%%%%%%%%%%%%%%%%%%%%%%%%%%%%%%%%%%%%%%

\begin{figure}[ht]
\caption{Transformation rules for eliminating transitivity\label{fig-trans}}
\begin{trans}
(a) & Cx & R(Lx)\\
(b) & R(Rx) & Rx\\
(c) & L(Lx) & Lx\\
(d) & L(Rx) & R(Lx)\\[2mm]

(e) & TIx & x\\[2mm]

(f) & T(Sx)y & S(Txy)\\[2mm]

(g) & T(Lx)y & L(Txy)\\[2mm]

(h) & T(RI)x & Lx\\
(i) & T(R(Sx))y & S(T(Rx)y)\\
(j) & T(R(Lx))y & L(T(Rx)y)\\
(k) & T(R(Pxy))I & R(Pxy)\\
(l) & T(R(Pxy))(Sz) & \bot\\
(m) & T(R(Pxy))(Lz) & T(Pxy)(Lz)\\
(n) & T(R(Pxy))(Rz) & R(T(R(Pxy))z)\\
(p) & T(R(Pxy))(Pzt) & P(Tz(Lx))(Ty(Lt))\\[2mm]

(q) & T(Pxy)I & Pxy\\
(r) & T(Pxy)(Sz) & \bot\\
(s) & T(Pxy)(LI) & R(Pxy)\\
(t) & T(Pxy)(L(Sz)) & \bot\\
(u) & T(Pxy)(L(Pzt)) & P(Tz(Lx))(Ty(Lt))\\
(v) & T(Pxy)(Rz) & R(T(Pxy)z)\\
(w) & T(Pxy)(Pzt) & P(Tzx)(Tyt)\\[2mm]

(1) & S\bot & \bot\\
(2) & L\bot & \bot\\
(3) & R\bot & \bot\\
(4) & P\bot x & \bot\\
(5) & Px\bot & \bot\\
(6) & T\bot x & \bot\\
(7) & Tx\bot & \bot\\
\end{trans}

Some of these rules are particular instances of the following more
general transformations:

\begin{trans}
(k') (q') & TxI & x\\
(n') (v') & Tx(Ry) & R(Txy)\\
(m') & T(Rx)(Ly) & Tx(Ly)\\
(s') & Tx(LI) & Rx\\
\end{trans}
\end{figure}

%%%%%%%%%%%%%%%%%%%%%%%%%%%%%%%%%%%%%%%%%%%%%%%%%%%%%%%%%%%%%%%%%%%%%%%%%%%%%%

\begin{lst}{}
\item [\bf(a)] $Cx\a R(Lx)$

\begin{ded}
T\ad T'\quad T'\le U'\quad U'\ad U
\justifies T\le U \using C
\end{ded}

\noindent
can be transformed into:

\begin{ded}
\[T\ad T'\quad T'\le U'
\justifies T\le U'\using L\]\quad U'\ad U
\justifies T\le U\using R
\end{ded}

%%%%%%%%%%%%%%%%%%%%%%%%%%%%%%%%%%%%%%%%%%%%%%%%%%%%%%%%%%%%%%%%%%%%%%%%%%%%%%

\item [\bf(b)] $R(Rx) \a Rx$

\begin{ded}
\[T\le U'\quad U'\ad U
\justifies T\le U\using R\]\quad U\ad U''
\justifies T\le U''\using R
\end{ded}

\noindent
can be transformed into:

\begin{ded}
T\le U'\quad U'\ad U''
\justifies T\le U''\using R
\end{ded}

\noindent
by confluence of $\a$.

%%%%%%%%%%%%%%%%%%%%%%%%%%%%%%%%%%%%%%%%%%%%%%%%%%%%%%%%%%%%%%%%%%%%%%%%%%%%%%

\item [\bf(c)] $L(Lx) \a Lx$

Like (b).

%%%%%%%%%%%%%%%%%%%%%%%%%%%%%%%%%%%%%%%%%%%%%%%%%%%%%%%%%%%%%%%%%%%%%%%%%%%%%%

\item [\bf(d)] $L(Rx) \a R(Lx)$

\begin{ded}
T\ad T'\quad \[T'\le U'\quad U'\ad U
             \justifies T'\le U\using R\]
\justifies T\le U\using L
\end{ded}

\noindent
can be transformed into:

\begin{ded}
\[T\ad T'\quad T'\le U'
\justifies T\le U'\using L\]\quad U'\ad U
\justifies T\le U\using R
\end{ded}

Note that the inverse transformation $R(Lx)\a L(Rx)$ is valid too.

%%%%%%%%%%%%%%%%%%%%%%%%%%%%%%%%%%%%%%%%%%%%%%%%%%%%%%%%%%%%%%%%%%%%%%%%%%%%%%

\item [\bf(e)] $TIx \a x$

\begin{ded}
\[\justifies T\le T\using I\]\quad T\le U
\justifies T\le U\using T
\end{ded}

\noindent
can be transformed into:

\[T\le U\]

%%%%%%%%%%%%%%%%%%%%%%%%%%%%%%%%%%%%%%%%%%%%%%%%%%%%%%%%%%%%%%%%%%%%%%%%%%%%%%

\item [\bf(f)] $T(Sx)y \a S(Txy)$

\begin{ded}
\[C^b\vt\le T
\justifies C^a\vt\le T\using S\]\quad T\le U
\justifies C^a\vt\le U\using T
\end{ded}

\noindent
can be transformed into:

\begin{ded}
\[C^b\vt\le T\quad T\le U
\justifies C^b\vt\le U\using T\]
\justifies C^a\vt\le U\using S
\end{ded}

%%%%%%%%%%%%%%%%%%%%%%%%%%%%%%%%%%%%%%%%%%%%%%%%%%%%%%%%%%%%%%%%%%%%%%%%%%%%%%

\item [\bf(g)] $T(Lx)y \a L(Txy)$

\begin{ded}
\[T\ad T'\quad T'\le U
\justifies T\le U\using L\]\quad U\le V
\justifies T\le V\using T
\end{ded}

\noindent
can be transformed into:

\begin{ded}
T\ad T'\quad \[T'\le U\quad U\le V
             \justifies T'\le V\using T\]
\justifies T\le V\using L
\end{ded}

%%%%%%%%%%%%%%%%%%%%%%%%%%%%%%%%%%%%%%%%%%%%%%%%%%%%%%%%%%%%%%%%%%%%%%%%%%%%%%

\item [\bf(h)] $T(RI)x \a Lx$

\begin{ded}
\[\[\justifies T\le T\using I\]\quad T\ad T'
          \justifies T\le T'\using R\]             \quad T'\le U
                    \justifies T\le U\using T
\end{ded}

\noindent
can be transformed into:

\begin{ded}
T\ad T'\quad T'\le U
\justifies T\le U\using L
\end{ded}

%%%%%%%%%%%%%%%%%%%%%%%%%%%%%%%%%%%%%%%%%%%%%%%%%%%%%%%%%%%%%%%%%%%%%%%%%%%%%%

\item [\bf(i)] $T(R(Sx))y \a S(T(Rx)y)$

\begin{ded}
\[\[C^b\vt\le T
\justifies C^a\vt\le T\using S\]\quad T\ad T'
\justifies C^a\vt\le T'\using R\]\quad T'\le U
\justifies C^a\vt\le U\using T
\end{ded}

\noindent
can be transformed into:

\begin{ded}
\[\[C^b\vt\le T\quad T\ad T'
\justifies C^b\vt\le T'\using R\] \quad T'\le U
\justifies C^b\vt\le U\using T\]
\justifies C^a\vt\le U\using S
\end{ded}

%%%%%%%%%%%%%%%%%%%%%%%%%%%%%%%%%%%%%%%%%%%%%%%%%%%%%%%%%%%%%%%%%%%%%%%%%%%%%%

\item [\bf(j)] $T(R(Lx))y \a L(T(Rx)y)$

By combination of (g) and the inverse of (d).

%%%%%%%%%%%%%%%%%%%%%%%%%%%%%%%%%%%%%%%%%%%%%%%%%%%%%%%%%%%%%%%%%%%%%%%%%%%%%%

\item [\bf(k')] $TxI \a x$

Like (e).

%%%%%%%%%%%%%%%%%%%%%%%%%%%%%%%%%%%%%%%%%%%%%%%%%%%%%%%%%%%%%%%%%%%%%%%%%%%%%%

\item [\bf(l)] $T(R(Pxy))(Sz) \a \bot$

\begin{ded}
\[\[U'\le U\quad V\le V'
\justifies (x:U)V\le (x:U')V'\using P\]\quad (x:U')V'\ad C^a\vt
\justifies (x:U)V\le C^a\vt\using R\]\quad
\[C^b\vt\le T\justifies C^a\vt\le T\using S\]
\justifies (x:U)V\le T\using T
\end{ded}

\noindent
is not possible since $(x:U')V'$ and $C^a\vt$ have no common reduct
since $C$ is constant.

%%%%%%%%%%%%%%%%%%%%%%%%%%%%%%%%%%%%%%%%%%%%%%%%%%%%%%%%%%%%%%%%%%%%%%%%%%%%%%

\item [\bf(n')] $Tx(Ry) \a R(Txy)$

\begin{ded}
T\le U\quad \[U\le V'\quad V'\ad V
            \justifies U\le V\using R\]
\justifies T\le V\using T
\end{ded}

\noindent
can be transformed into:

\begin{ded}
\[T\le U\quad U\le V'\justifies T\le V'\using T\]
\quad V'\ad V\justifies T\le V\using R
\end{ded}

%%%%%%%%%%%%%%%%%%%%%%%%%%%%%%%%%%%%%%%%%%%%%%%%%%%%%%%%%%%%%%%%%%%%%%%%%%%%%%

\item [\bf(m')] $T(Rx)(Ly) \a Tx(Ly)$

\begin{ded}
\[T\le U\quad U\ad U'\justifies T\le U'\using R\]
\quad \[U'\ad U''\quad U''\le V\justifies U'\le V\using L\]
\justifies T\le V\using T
\end{ded}

\noindent
can be transformed into:

\begin{ded}
T\le U\quad \[U\ad U''\quad U''\le V\justifies U\le V\using L\]
\justifies T\le V\using T
\end{ded}

\noindent
by confluence of $\a$.

%%%%%%%%%%%%%%%%%%%%%%%%%%%%%%%%%%%%%%%%%%%%%%%%%%%%%%%%%%%%%%%%%%%%%%%%%%%%%%

\item [\bf(p)] $T(R(Pxy))(Pzt) \a P(Tz(Lx))(Ty(Lt))$

\begin{ded}
\[\[U_2\le U_1\quad V_1\le V_2\justifies (x:U_1)V_1\le (x:U_2)V_2\using P\]
\quad (x:U_2)V_2\ad (x:U_3)V_3\justifies (x:U_1)V_1\le (x:U_3)V_3\using R\]
\quad \[U_4\le U_3\quad V_3\le V_4\justifies (x:U_3)V_3\le (x:U_4)V_4\using P\]
\justifies (x:U_1)V_1\le (x:U_4)V_4\using T
\end{ded}

\noindent
can be transformed into:

\begin{ded}
\[U_4\le U_3\quad \[U_3\ad U_2\quad U_2\le U_1\justifies U_3\le U_1\using L\]
\justifies U_4\le U_1\using T\]\quad
\[V_1\le V_2\quad \[V_2\ad V_3\quad V_3\le V_4\justifies V_2\le V_4\using L\]
\justifies V_1\le V_4\using T\]
\justifies (x:U_1)V_1\le (x:U_4)V_4\using P
\end{ded}

%%%%%%%%%%%%%%%%%%%%%%%%%%%%%%%%%%%%%%%%%%%%%%%%%%%%%%%%%%%%%%%%%%%%%%%%%%%%%%

\item [\bf(r)] $T(Pxy)(Sz) \a \bot$

Like (l).

%%%%%%%%%%%%%%%%%%%%%%%%%%%%%%%%%%%%%%%%%%%%%%%%%%%%%%%%%%%%%%%%%%%%%%%%%%%%%%

\item [\bf(s')] $Tx(LI) \a Rx$

Like (h).

%%%%%%%%%%%%%%%%%%%%%%%%%%%%%%%%%%%%%%%%%%%%%%%%%%%%%%%%%%%%%%%%%%%%%%%%%%%%%%

\item [\bf(t)] $T(Pxy)(L(Sz)) \a \bot$

Like (l).

%%%%%%%%%%%%%%%%%%%%%%%%%%%%%%%%%%%%%%%%%%%%%%%%%%%%%%%%%%%%%%%%%%%%%%%%%%%%%%

\item [\bf(u)] $T(Pxy)(L(Pzt)) \a P(Tz(Lx))(Ty(Lt))$

Like (p).

%%%%%%%%%%%%%%%%%%%%%%%%%%%%%%%%%%%%%%%%%%%%%%%%%%%%%%%%%%%%%%%%%%%%%%%%%%%%%%

\item [\bf(w)] $T(Pxy)(Pzt) \a P(Tzx)(Tyt)$

Like (p).

\end{lst}

The above rules form a terminating rewrite system. For $L$ and $R$,
the recursive calls are strictly smaller (take $L<R$). For $Tuv$, the
measure $(|u|+|v|,|v|)$, where $|u|$ is the size of $u$, strictly
decreases lexicographically. Now, it is easy to see that $T$ occurs in
no normal form of $Tuv$ if $u$ and $v$ are closed terms ($T$ is
completely defined). We proceed by induction on the measure. The only
undefined cases for $T$ are $T(R(Pxy))(Tzt)$, $T(Pxy)(L(Tzt))$,
$T(Pxy)(Tzt)$ and $T(Txy)z$. By induction hypothesis, $T$ occurs in no
normal form of $Tzt$ or $Txy$. Therefore, we fall in the defined cases
and we can conclude by induction hypothesis.

%% file: exp.tex
%%%%%%%%%%%%%%%%%%%%%%%%%%%%%%%%%%%%%%%%%%%%%%%%%%%%%%%%%%%%%%%%%%%%%%%%%%%%%%
% expansion elimination
%%%%%%%%%%%%%%%%%%%%%%%%%%%%%%%%%%%%%%%%%%%%%%%%%%%%%%%%%%%%%%%%%%%%%%%%%%%%%%

\section{Expansion elimination}
\label{sec-exp}

In this section, we prove Theorem \ref{thm-exp-elim} by following
Chen's technique \cite{chen98thesis}. We introduce the following term
algebra for representing the subtyping deductions:

\begin{center}
$d ::= I ~|~ S ~|~ Ed ~|~ Rd ~|~ Pdd$
\end{center}

\noindent
where $\bot$ stands for some impossible case, $I$ for (refl), $S$ for
(symb), $C$ for (conv), $E$ for (exp), $R$ for (red), and $P$ for
(prod).

We now prove that the following transformation rules are valid, that
is, a deduction matching a left hand-side can be replaced by the
corresponding right hand-side.

\begin{trans}
(a) & E(Rx) & R(Ex)\\
(b) & E(Pxy) & P(Ex)(Ey)\\
(c) & EI & RI\\
(d) & ES & RS\\
(e) & E(Ex) & Ex\\
\end{trans}

\begin{bfenumalphai}
% E(Rx)\a R(Ex)
\item $E(Rx)\a R(Ex)$

Assume that we have the following deduction:

\begin{ded}
\[T'\a^* T''\le U''\als U'
\justifies T\als T'\le U'\a^* U\using R\]
\justifies T\le U\using E
\end{ded}

By confluence, there exist $T'''$ and $U'''$ such that $T\a^* T'''\als
T''$ and $U\a^* U'''\als U''$. So, the deduction can be transformed
into:

\begin{ded}
\[T'''\als T''\le U''\a^* U'''
\justifies T\a^* T'''\le U'''\als U\using E\]
\justifies T\le U\using R
\end{ded}

% E(Pxy)\a P(Ex)(Ey)
\item $E(Pxy)\a P(Ex)(Ey)$

Assume that we have the following deduction:

\begin{ded}
\[C\le A\quad B\le D
\justifies T\als (x:A)B\le (x:C)D\a^* U\using P\]
\justifies T\le U\using E
\end{ded}

Then, $T=(x:A')B'$ with $A\a^* A'$ and $B\a^* B'$, and $U=(x:C')D'$
with $C\a^* C'$ and $D\a^* D'$. So, the deduction can be transformed
into:

\begin{ded}
\[C'\als C\le A\a^* A'\justifies C'\le A'\using E\]
\quad \[B'\als B\le D\a^* D'\justifies B'\le D'\using E\]
\justifies T\le U\using P
\end{ded}

% EI\a RI
\item $EI\a RI$

By confluence, as in (a) but with $T'=T''=U''=U'$.

% ES\a
\item $ES\a RS$

Assume that we have the following deduction:

\begin{ded}
\[a\le_\cA b\justifies T\als C^a\vt\le C^b\vt\a^* U\using S\]
\justifies T\le U\using E
\end{ded}

Then, $T=C^a\vu$ with $\vt\a^*\vu$ and $U=C^b\vv$ with
$\vt\a^*\vv$. By confluence, there exists $\vw$ such that
$\vu\a^*\vw\als\vv$. So, the deduction can be transformed into:

\begin{ded}
\[a\le_\cA b\justifies T\a^* C^a\vw\le C^b\vw\als U\using S\]
\justifies T\le U\using R
\end{ded}

% E(Ex)\a Ex
\item $E(Ex)\a Ex$

Immediate.
\end{bfenumalphai}

Now, the rewrite system defined by these transformation rules is
clearly terminating and confluent (there is no critical pair). Since
it defines $E$ completely, no normal form of a closed term may contain
$E$.